\documentclass[a4paper,11pt]{article}
\pdfoutput=1
\usepackage[T1]{fontenc}
\usepackage[english]{babel}
\usepackage{amsthm,amsfonts}
\usepackage{empheq}
\usepackage{dsfont}
\hyphenchar\font=\string"7F

\usepackage{breqn}
\usepackage{jheppub}

\usepackage{tikz}
\usetikzlibrary{arrows,matrix,calc}
\tikzset{>=stealth'}
\tikzstyle{shiftup}=[transform canvas={yshift=0.25em}]
\tikzstyle{shiftdn}=[transform canvas={yshift=-0.25em}]
\tikzstyle point=[minimum size=1mm,inner sep=0pt,outer sep=0pt,shape=circle,fill=black]

\makeatletter
\renewcommand{\@dotsep}{10000}
\makeatother

\DeclareMathOperator{\Tr}{Tr}
\newcommand{\DFTwzw}{DFT${}_{\mathrm{WZW}}$}

\title{\huge\boldmath The Topology of Double Field Theory}
\preprint{}

\author[a]{Falk Hassler}
\emailAdd{fhassler@unc.edu}

\affiliation[]{University of North Carolina\\Department of Physics and Astronomy\\
Phillips Hall, CB \#3255, 120 E. Cameron Ave., Chapel Hill, NC 27599-3255, USA}

\abstract{We describe the doubled space of Double Field Theory as a group manifold $G$ with an arbitrary generalized metric. Local information from the latter is not relevant to our discussion and so $G$ only captures the topology of the doubled space. Strong Constraint solutions are maximal isotropic submanifold $M$ in $G$. We construct them and their Generalized Geometry in Double Field Theory on Group Manifolds. In general, $G$ admits different physical subspace $M$ which are Poisson-Lie T-dual to each other. By studying two examples, we reproduce the topology changes induced by T-duality with non-trivial $H$-flux which were discussed by Bouwknegt, Evslin and Mathai \cite{Bouwknegt:2003vb}.} 

\begin{document}
\maketitle

\section{Introduction and Summary}
Dualities play a crucial role in our current understanding of string theory. S- and T-duality connect the five perturbative superstring theories in an intriguing way and led to the discovery of M-theory \cite{Hull:1994ys,Witten:1995ex}. Due their complexity, usually only their low-energy effective actions are studied. Unfortunately, T-duality is not directly manifest at this level. In the NS/NS sector of type IIA/B on a $d$-dimensional torus, it is mediated by the Buscher rules \cite{Buscher:1987sk}. They link dual target spaces in complicated, non-linear way.

The major objective of Double Field Theory\footnote{Based on a superspace approach, a target space theory with manifest SO($d,d+n$) symmetry was worked out by Siegel \cite{Siegel:1993th}. Independently Hull and Zwiebach derived the low energy effective action and gauge transformation of DFT from Closed String Field Theory (CSFT) \cite{Hull:2009mi,Hull:2009zb,Hohm:2010pp}. Both theories have the same essential ingredients.} \cite{Siegel:1993th,Hull:2009mi,Hull:2009zb,Hohm:2010pp} (for reviews see \cite{Aldazabal:2013sca,Hohm:2013bwa}) is to make the T-duality group O($d,d,\mathbb{Z}$) of a $d$-dimensional torus manifest. To this end, the theory is formulated on a doubled space with momentum and winding coordinates. There is a clear distinction between weakly and strongly constrained DFT. While the former is only valid on a tours, the latter works for arbitrary target spaces. It imposes the Strong Constraint (SC) which restricts all fields to depend on a $d$-dimensional physical space only. In doing so all supergravity backgrounds are accessible in DFT. At the same time the notion of the doubled space becomes more subtle. Depending on the target space topology there are in general much less winding modes than on a torus. Hence, one is inclined to ask: What is the significance of the doubled space in the strongly constraint theory? Does it still captures relevant information about T-duality or is it just a convenient bookkeeping device?

Currently, the doubled space of DFT is argued not to be a differentiable manifold \cite{Hohm:2013bwa}\footnote{This statement holds for the current standard formulation of DFT introduced by Hohm, Hull and Zwiebach. There are also other approaches, like for example \cite{Vaisman:2012ke,Vaisman:2012px,Cederwall:2014kxa,Cederwall:2014opa} which treat the doubled space as a differentiable manifold.}. In a single patch this does not cause problems because one is always able to solve the SC by choosing $d$ out of the $2d$ coordinates for the physical space. All fields in the theory only depend on them. Gluing different patches together is more involved. Finite generalized diffeomorphisms are the natural candidates for transition functions and therefore were intensively study recently \cite{Hohm:2012gk,Park:2013mpa,Berman:2014jba,Hull:2014mxa,Naseer:2015tia}. But they rely on solving the SC first and patching afterwards which results in subtleties for target spaces with $H$-flux in a non-trivial cohomology class \cite{Papadopoulos:2014mxa,Howe:2016ggg}. Hence going beyond toroidal backgrounds, the global structure of the doubled space is not fully understood yet. Especially in the context of T-duality this is a pity. In a coarse approximation, it exchanges momentum and winding modes which crucially depend on the topology of the target space. Thus, global properties have to play a central role. Indeed, T-duality can induce topology changes of the target space \cite{Alvarez:1993qi,Bouwknegt:2003vb,Hull:2006qs} which has important implications for brane and string charges. This paper presents a different approach to the doubled space. We describe it as a $2d$-dimensional Lie group $G$ which admits an embedding into O($d$,$d$). Doubled geometries of this kind \cite{Hull:2005hk,Hull:2007jy,Dall'Agata:2008qz,Hull:2009sg} were originally introduced by Hull and Reid-Edwards. They have a natural interpretation in lower dimensional gauged supergravities which arise from Scherk-Schwarz compactifications \cite{Scherk:1979zr,Scherk:1978ta,Aldazabal:2011nj,Geissbuhler:2011mx,Geissbuhler:2013uka}. However, they did not affect the perspective on the doubled space in DFT until recently when \DFTwzw{} \cite{Blumenhagen:2014gva,Blumenhagen:2015zma,Bosque:2015jda} was proposed. It is a formulation which is derived from a Wess-Zumino-Witten model and implements a non-trivial $H$-flux by construction. Like in DFT, the doubled space $G$ is equipped with an arbitrary generalized metric. For the following discussions this metric is irrelevant. Only the topology of $G$ matters. By solving the SC of \DFTwzw, one embeds the theory's physical subspace $M$, which is restricted to have the topology of a coset space, in $G$. In general, there are different SC solutions and each of them gives rise to a dual background. Two explicit examples are studied in this paper. They reproduce the topology changes presented in \cite{Bouwknegt:2003vb}. Furthermore, this approach allows to patch the doubled space with standard diffeomorphisms instead of the finite generalized ones mentioned above. It is known that the latter are insufficient in implementing the full O($d$,$d$) and more general transformations are required. Normally, they are introduced by hand. But they also admit a geometric interpretation \cite{Cederwall:2014opa}. Our approach exploits this geometric perspective and so circumvents the problems outlined above. 

A SC solution identifies a $d$-dimensional physical subspace $M$ in $G$. To obtain it, we choose a maximal isotropic subgroup $H\subset G$ and consider the coset $M=G/H$. In general one can choose between different subgroups. Each one gives rise to a T-dual solution. The same prescription is used to define Poisson-Lie T-duality \cite{Klimcik:1995ux,Klimcik:1996nq,Severa:2016prq}. Of course T-duality also imposes additional constraints on the generalized metric and dilaton. We do not discuss these restrictions in this paper, because we are merely interested in metric independent statements. Furthermore, the notion of T-duality used here is Poisson-Lie T-duality \cite{Klimcik:1995ux}. It includes abelian and non-abelian T-duality as special cases and is currently the most general notion of T-duality available. There might be the concern that Poisson-Lie T-duality is not a full symmetry of string theory. In contrast to abelian T-duality, it does not hold to all orders in the string coupling $g_s$ and the string tension $\alpha'$. Still, at the two derivative level (leading order $\alpha'$) on which we are discussing DFT here, it has exactly the same properties as abelian T-duality. In particular already the canonical example of the T-duality chain for the torus with $H$-flux, which we present in section~\ref{sec:torusHflux}, is an example of Poisson-Lie T-duality bases on the non-abelian Drinfeld double $\mathfrak{cso}(1,0,3)$.

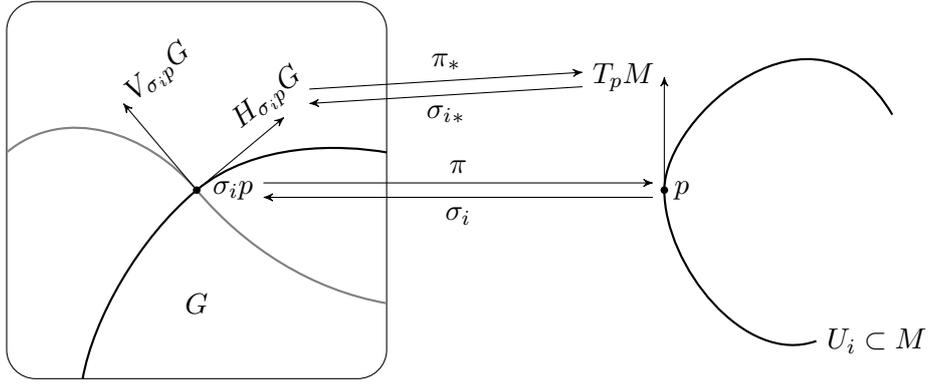
\begin{figure}
	\centering
  \begin{tikzpicture}
		\begin{scope}
	    \draw[rounded corners=1em] (-1, -1) rectangle (4, 4);
      \draw[thick] (0, -1) .. controls ++(80:0.8) and ++(-140:1) .. (1.5, 1.5) .. controls ++(40:1) and ++(170:0.6) .. (4, 2);
			\draw[gray,thick] (4, 0) .. controls ++(170:1) and ++(-50:1) .. (1.5, 1.5) .. controls ++(130:1) and ++(40:1) .. (-1, 2);
      \draw[->] (1.5,1.5) node[point] {} -- ++(40:1.5) node[anchor=south,rotate=40,name=HG] {$H_{\sigma_i p} G$};
      \draw[->] (1.5,1.5) node[anchor=west,xshift=0.2em,name=sigmaP] {$\sigma_i p$} -- ++(130:1.5) node[anchor=west,rotate=40,name=VG] {$V_{\sigma_i p} G$};
			\node[at={(1.5, 0)},anchor=center] {$G$};
		\end{scope}
		\begin{scope}[xshift=20em]
      \draw[thick] (2,-0.5) node[anchor=west] {$U_i \subset M$} .. controls ++(200:1) and ++(-90:1) .. (0, 1.5) .. controls ++(90:1) and ++(120:2) .. (3, 2.5);
			\draw[->] (0,1.5) node[point] {} -- ++(90:1.5) node[anchor=east,name=TM] {$T_p M$};
			\node[at={(0,1.5)},anchor=west,name=P] {$p$};
		\end{scope}
    \draw[->,shiftdn] (TM) -- (HG) node[midway, below] {${\sigma_i}_*$};
    \draw[<-,shiftup] (TM) -- (HG) node[midway, above] {$\pi_*$};
    \draw[->,shiftdn] ($(P)-(1em,0)$) -- (sigmaP) node[midway, below] {$\sigma_i$};
    \draw[<-,shiftup] ($(P)-(1em,0)$) -- (sigmaP) node[midway, above] {$\pi$};
  \end{tikzpicture}
  \caption{The curved, black line represents all directions in $G$ which solve the SC. Their tangent space forms the horizontal subspace defined by the connection one-from $\omega$ as $\omega(X)=0 \,\, \forall X\in H G$. All tangent vectors on the physical space $M$ are mapped to $H G$ by the differential map ${\sigma_i}_*$ in the appropriate patch $U_i \subset M$.}\label{fig:maps}
\end{figure}

Only fixing $H$ is not sufficient to construct a complete SC solution. Additionally, we need a map $\sigma_i: U_i \rightarrow G$ in each patch $U_i \subset M$. Fixing this map is the main challenge we have to face. To this end, $G$ is interpreted as a $H$-principal bundle $\pi: G \rightarrow M$ with a connection one-form $\omega$ splitting its tangent space into a horizontal and a vertical part $T G = H G \oplus V G$. We choose $\omega$ in a particular way such that $H G$ contains all tangent vectors of the $d$ directions in $G$ solving the SC. Finally, we have to choose a $\sigma_i$ whose differential map only has $H G$ as image, but not $V G$. Equivalently, we have to fulfill the constraint
\begin{equation}
  \omega {\sigma_i}_* = 0 = \sigma_i^* \omega = A_i = 0
\end{equation}
which requires a locally vanishing gauge potential $A_i$ on $M$. This procedure fixes both $\sigma_i$ and $\omega$. First, we choose $\omega$ such that the field strength for all $A_i$ in all patches vanishes. In this case the gauge potential is locally a pure gauge and we can set it locally to zero by an appropriate gauge transformation, fixing $\sigma_i$ at the same time. If the bundle is not trivial, we can not set $A_i$ to zero globally. However as we explain in section~\ref{sec:HPrincipal}, this is not required to find a global solution of the SC. It is totally sufficient to have a flat connection. We further show that the connection one-form is isomorphic to a pure spinor which specifies the polarization on the group manifold. Figure \ref{fig:maps} illustrates the relation between $G$ and $M$.
All remaining unphysical directions are represented by the gray line in this figure. Their tangent vectors span $V G$ and are isomorphic to elements in the Lie algebra $\mathfrak{h}$ of $H$. By defining an isomorphism $\eta: \mathfrak{h} \rightarrow T^* M$, we are able to identify $V_{\sigma_i p} G $ with the cotangent space $T^*_p M$ at each point $p\in U_i$. Finally, we obtain the exact sequence
\begin{equation}
\begin{tikzpicture}\label{eqn:exactCourantAlg2}
  \matrix (m) [matrix of math nodes, column sep=3em] {%
    0 & T^* M & T G & T  M & 0 \\ };
  \draw[->,shiftup] (m-1-1.east) -- (m-1-2.west |- m-1-1);
  \draw[<-,shiftdn] (m-1-1.east) -- (m-1-2.west |- m-1-1);
  \draw[->,shiftup] (m-1-2.east) -- (m-1-3.west |- m-1-2) node[midway, above] {$(\eta^{-1})^\sharp$};
  \draw[<-,shiftdn] (m-1-2.east) -- (m-1-3.west |- m-1-2) node[midway, below] {$\eta \omega$};
  \draw[->,shiftup] (m-1-3.east) -- (m-1-4.west |- m-1-3) node[midway, above] {$\pi_*$};
  \draw[<-,shiftdn] (m-1-3.east) -- (m-1-4.west |- m-1-3) node[midway, below] {${\sigma_i}_*$};
  \draw[->,shiftup] (m-1-4.east) -- (m-1-5.west |- m-1-4);
  \draw[<-,shiftdn] (m-1-4.east) -- (m-1-5.west |- m-1-4);
\end{tikzpicture}
\end{equation}
where ${}^\sharp$ assigns a fundamental vector field to each element of $\mathfrak{h}$. In conjunction with the generalized Lie derivative of \DFTwzw, this sequence represents an exact Courant algebroid. We use its maps to define a generalized frame field $\hat E_A$. It is an element of O($d$,$d$) and allows to pull all fields on the group manifold to the generalized tangent space $T^* M \oplus T M$. Thus, for each solution of the SC we obtain a Generalized Geometry (GG) \cite{Hitchin:2004ut,Gualtieri:2003dx} with a twisted Courant bracket. There are two ways how such a twist shows up \cite{Koerber:2010bx}. It can be realized as a twist term in the Courant bracket or as a gerbe for the $B$-field in the frame field $\hat E_A$. In the latter case, the generalized tangent space is replaced by a twisted bundle $E$ which only looks locally like the sum of $M$'s tangent and co-tangent bundle. In general SC solutions in \DFTwzw{} have both contributions. How they are distributed depends on the specific properties of the group manifold $G$ and the chosen subgroup $H$.

We present two explicit examples, the torus with $H$-flux and the three-sphere $S^3$ with $H$-flux. They implement T-duality in very different ways. For the former, we obtain the T-duality chain \cite{Shelton:2005cf,Dabholkar:2005ve}
\begin{equation}
  H_{ijk} \stackrel{\mathcal{T}_i}{\longrightarrow} f^{i}_{jk}
  \stackrel{\mathcal{T}_j}{\longrightarrow} Q_{k}^{ij}
\end{equation}
($\mathcal{T}_i$ denotes T-duality along $x^i$) by choosing different, maximal isotropic subgroups $H \subset G$. All possible subgroups are classify by Lie algebra cohomology and related by O($d$)$\times$O($d$) transformations. Similar results are known in the current formulation of DFT. One important difference is that the $H$-flux of the background splits into a gerbe and a twist contribution. This splitting is not arbitrary but completely fixed by the group manifold $G$=CSO($1$,$0$,$3$). At the same time the solution of the SC is twisted. While in each patch one can split the $2d$ coordinates on $G$ into $d$ physical and $d$ unphysical ones, they mix globally. T-duality acts as a change of coordinates on $G$. Only fields with an isometry along the T-duality direction solve the SC in both frames. We are not able to find SC solutions with $R$-flux.

For the $S^3$ with $\mathbf{h}$ units of $H$-flux the only possible subgroup of $G$=SO($4$) is $H=$SO($3$). As a simple Lie group its algebra $\mathfrak{h}$ does not admit any non-trivial deformations. Its SC solution implements the complete $H$-flux in the twist of the Courant bracket. Despite the lack of different subgroups, one can still obtain a T-dual background by modding out a discrete subgroup which results in $M$=$\mathbb{Z}_{\mathbf{h}}\setminus$SO($4$)$/$SO($3$), the lens space L$(\mathbf{h},1)$ with one unit of $H$-flux, as physical space. However, we do not obtain a $T^2$ fibration over $S^2$ as described in \cite{Schulz:2011ye}. The reason is that the CFT for the dual target space is obtained from a SU($2$) WZW-model at level $\mathbf{h}$ by orbifolding a finite group \cite{Giddings:1993wn,Maldacena:2001ky}. As result there is only a finite number of winding modes in the twisted sector. They are not sufficient to furnish all Fourier modes on a circle and thus cannot be associated to the winding coordinate of a $S^1$ dual to the one of the $S^1$ Hopf fiber of $S^3$.

All results obtained in this work allow to refine the statements about the status of \DFTwzw{} compared to the traditional formulation. Locally, they are equivalent after solving the SC \cite{Blumenhagen:2015zma,Hohm:2015ugy}. Globally, \DFTwzw{} contains additional information about the topology of the doubled space. The remaining part of the paper is organized as follows: A compact review of the salient features of \DFTwzw{} is given in section~\ref{sec:review}. The technique to solve the SC, we already outlined above, is developed in section~\ref{sec:solSC}. For symmetric spaces $M$, we show how such solutions can be constructed very explicitly. Section~\ref{sec:examples} presents the tours and the $S^3$, both with $\mathbf{h}$ units of $H$-flux, as examples. We close with an outlook in section~\ref{sec:outlook} which highlights possible applications and directions for future projects.

\section{Review of Double Field Theory on Group Manifold}\label{sec:review}
In the following we review the features of \DFTwzw{} \cite{Blumenhagen:2014gva,Blumenhagen:2015zma} which are essential to discuss non-trivial solutions of the SC. A broader review is given in \cite{Hassler:2015pea}. The theory is formulated on a doubled space equivalent to a Lie group $G$ with the coordinates $X^I$ ($I=1,\dots,2 d$). $G$ carries an O($d,d$) structure which allows to introduce a globally defined metric $\eta_{IJ}$ with split signature. Because Lie groups are parallelizable, one is able to introduce a global generalized frame $E_A{}^I\in$GL($2d$) (and its inverse transpose $E^A{}_I$) which is called background generalized vielbein. It is used to bring the $\eta$-metric in the canonical form
\begin{equation}
  E_A{}^I \eta_{IJ} E_B{}^J = \eta_{AB} = \begin{pmatrix}
    0 & \delta_b^a \\ \delta_a^b & 0
  \end{pmatrix}\,.
\end{equation}
In general there are different ways how to choose this canonical form. They are all related by GL($2d$) transformations. In \cite{Blumenhagen:2014gva}, it is diagonal and makes the splitting in left- and right-movers on the string worldsheet manifest. Here, we use a representation inspired by the standard formulation of DFT. Thus, flat indices $A$ have the structure
\begin{equation}\label{eqn:gen&dual}
  t_A = \begin{pmatrix} t^a & t_a \end{pmatrix} \quad \text{and} \quad
  \theta^A = \begin{pmatrix} \theta_a & \theta^a \end{pmatrix}
\end{equation}
where $a=1,\dots,d$. We denote the generators of $G$'s Lie algebra $\mathfrak{g}$ by $t_A$ and $\theta^A$ is the corresponding dual one-form ($\theta^A (t_B) = \delta^A_B$). A particular choice for $\mathfrak{g}$ is picked by fixing the structure coefficients
\begin{equation}
  [t_A, t_B] = F_{AB}{}^C t_C\,.
\end{equation}
They implement non-trivial background fluxes in the theory. The Lie algebra structure carries over to the flat derivatives $D_A = E_A{}^I \partial_I$ which fulfill the analog identity
\begin{equation}
  [D_A, D_B] = F_{AB}{}^C D_C \quad\text{or equivalently} \quad
  F_{AB}{}^C = 2 D_{[A} E_{B]}{}^I E^C_I \,.
\end{equation}

\DFTwzw{} has two dynamical fields, the generalized metric $\mathcal{H}_{IJ}$ and the generalized dilaton $\phi$. Their dynamics is governed by the action \cite{Blumenhagen:2015zma}
\begin{equation}\label{eqn:action}
  S = \int d^{2d} X e^{-2 \phi} \, \mathcal{R}\,,
\end{equation}
which can be expressed in terms of the generalized curvature scalar
\begin{align}
  \mathcal{R} &= 4 \mathcal{H}^{IJ} \nabla_I \nabla_J \phi - \nabla_I \nabla_J \mathcal{H}^{IJ} - 4 \mathcal{H}^{IJ} \nabla_I \phi \, \nabla_J \phi + 4 \nabla_I \phi \, \nabla_J \mathcal{H}^{IJ} \nonumber \\ &\,+ \frac{1}{8} \mathcal{H}^{KL} \nabla_K \mathcal{H}_{IJ} \nabla_L \mathcal{H}^{IJ} - \frac{1}{2} \mathcal{H}^{IJ} \nabla_J \mathcal{H}^{LK} \nabla_K \mathcal{H}_{IL} + \frac{1}{6} F_{IKL} F_J{}^{KL} \mathcal{H}^{IJ}\,.
\label{eqn:gencurvature}
\end{align}
In this equation, a second covariant derivative $\nabla_A$ appears. It is connected to $D_A$ by the relation
\begin{equation}
  \nabla_A V^B = D_A V^B + \frac{1}{3} F^B{}_{AC} V^C\,.
\end{equation}
Furthermore, its action on curved indices is chosen such that it is compatible with the background generalized vielbein
\begin{equation}
  \nabla_A E_B{}^I = D_A E_B{}^I - \frac{1}{3} F^C{}_{AB} E_C{}^I + E_A{}^J \Gamma^I{}_{JK} E_B{}^K = 0\,,
\end{equation}
resulting in
\begin{equation}
  \nabla_I V^J = \partial_I V^J + \Gamma^J{}_{IK} V^K
    \quad \text{with} \quad
    \Gamma^I{}_{JK} = - \partial_J E_A{}^I E^A{}_K + \frac13 F^C{}_{AB} E_C{}^I E^A{}_J E^B{}_K\,.
\end{equation}
The \DFTwzw{} action \eqref{eqn:action} is invariant under generalized diffeomorphisms. Infinitesimally, they are generated by the generalized Lie derivative
\begin{equation}\label{eqn:genLie}
  \mathcal{L}_\xi V^I = \xi^J \nabla_J V^I +  \big( \nabla^I \xi_J - \nabla_J \xi^I \big) V^J\,.
\end{equation}
However, these transformations only close into a consistent algebra if the strong constraint (SC)
\begin{equation}\label{eqn:flatSC}
  D_A D^A \cdot = 0
\end{equation}
hold. $\cdot$ is a place holder for fields, parameters of generalized diffeomorphisms and arbitrary products of them. Basically, this constraint requires that all fields only depend on a $d$-dimensional subspace $M$ of the doubled space. Locally there is not more to add to this statement. Globally it is more subtle as this paper shows.

In addition to generalized diffeomorphisms, the action and the generalized Lie derivative transform covariantly under $2d$-diffeomorphisms. They are mediated by the standard Lie derivative \cite{Blumenhagen:2015zma}
\begin{equation}
  L_\xi V^I = \xi^J \partial_J V^I - V^J \partial_J \xi^I
\end{equation}
and their appearance seems natural since we started from a Lie group $G$ as doubled space. Remember, each Lie group is a differentiable manifold. Thus, we should be able to find all structures known from differential geometry and topology. We exploit them to construct explicit solutions of the SC in the next section. This is the main difference between \DFTwzw{} and the traditional formulation \cite{Hull:2009mi,Hohm:2010jy,Hohm:2010pp}. There, the doubled space does not carry any additional structures besides generalized diffeomorphisms.

\section{Solutions of the Section Condition}\label{sec:solSC}
For the consistency of the \DFTwzw{} the SC \eqref{eqn:flatSC} plays a crucial role. It can be trivially solved by making all fields and parameters of gauge transformations constant on the group manifold. Such solutions are used in the generalized Scherk-Schwarz compactifications presented in \cite{Bosque:2015jda}. Here, we are interested in a more general class. We construct explicit solutions by exploiting that the doubled space is a differentiable manifold which can be equipped with the structure of a $H$-principal bundle. In subsection~\ref{sec:HPrincipal}, we explain how such bundles capture all data that are required to describe a global solution of the SC by embedding a physical submanifold $M = G/H$ in $G$. While the generalized Lie derivative \eqref{eqn:genLie} acts on the tangent space $T G$ of the group manifold, we use the maps defining the $H$-principal bundle to pull it to the generalized tangent space $T^* M \oplus T M$. This link between DFT and GG is known since the early days \cite{Hull:2009zb}. However, we will see that globally the generalized tangent bundle gets twisted as a result of the non-trivial topology of the group manifold. Capturing this twist without any reference to the generalized metric is one of the merits of the formalism presented in this paper. As we finally discuss in subsection~\ref{sec:deformations}, the SC admits in general more than one solution. Each of them is invariant under arbitrary O($d,d$)/O($d$)$\times$O($d$) transformations. This statement is equivalent to having a free choice of a generalized metric $\mathcal{H}_{IJ}$ which lives on $M$. Hence, all statements in this paper are only sensitive to the topology of the group manifold. From this point of view \DFTwzw{} is as background independent as the traditional formulation \cite{Hohm:2010jy}. Additionally, there are also O($d$)$\times$O($d$) deformations which result in different subalgebras $H$ for the principal bundle. They are classified in terms of Lie algebra cohomology and, as we see in section~\ref{sec:examples}, produce different T-dual backgrounds.

\subsection{Reformulation as \texorpdfstring{$H$}{H}-Principal Bundle}\label{sec:HPrincipal}
We now present a systematic way to find SC solutions on a group manifolds. Following \cite{Berman:2012vc}, we first substitute its quadratic version through the equivalent linear constraint
\begin{equation}\label{eqn:sclinear}
  \Lambda^\alpha \Gamma_{\alpha\beta}^A D_A \cdot = 0\,.
\end{equation}
It allows us to specify a pure\footnote{A pure spinor is annihilated by the maximal number of linear independent $\Gamma$-matrices. Thus, it selects a maximally isotropic subspace as for example explain in section 3.3 of the review \cite{Koerber:2010bx}.}, Spin($d$,$d$) Majorana-Weyl (MW) spinor $\Lambda^\alpha$ for all distinct solutions. MW spinors carry indices $\alpha=1,\dots,2^{d-1}$ and $\Gamma_{\alpha\beta}^A$ denotes the corresponding $\Gamma$-matrices defined by
\begin{equation}\label{eqn:cliffordalg}
  \big\{\Gamma^A, \Gamma^B\big\} = 2 \eta^{AB}\,.
\end{equation}
We assign a value to this spinor on each points of the group manifold $G$. In order to relate different points, we remember that $G$ is equipped with a Lie algebra $\mathfrak{g}$ which generates infinitesimal translations. Its generators are completely specified by the structure coefficients $F_{AB}{}^C$ and can either act on the
\begin{equation}
  \text{vector} \quad  (t_A)_B{}^C = F_{AB}{}^C \quad \text{or MW spinor} \quad
  (t_A)^\alpha{}_\beta = \frac12 F_{ABC} (\Gamma^{BC})^{\alpha}{}_\beta 
\end{equation}
representation of O($d,d$). Applying the exponential map to them, one obtains group elements in these two representations denoted as $g_A{}^B$ and $g^\alpha{}_\beta$, respectively. Furthermore, assume we found a set of fields $f$ with a specific coordinate dependence such that they solve the linear constraint \eqref{eqn:sclinear} for a fixed $\Lambda^\alpha$. In this case there also exists another set of fields, we call them $f'$, with a different coordinate dependence
\begin{equation}
  D_A f' = (\mathrm{Ad}_g)_A{}^B D_B f
\end{equation}
solving the linear constraint for
\begin{equation}\label{eqn:shiftvag}
  {\Lambda'}^\alpha = g^\alpha{}_\beta \Lambda^\beta\,.
\end{equation}
We denote the adjoint action of a group element $g$ on a Lie algebra element in the first equation as
\begin{equation}
  (\mathrm{Ad}_g)_A{}^B t_B = g\, t_A g^{-1}\,.
\end{equation}
This property of the linear constraint \eqref{eqn:sclinear} is essential for what follows. It results from the fact that the $\Gamma$-matrices transform covariant under O($d,d$). By construction, $G$ is a subgroup of this group and thus \eqref{eqn:sclinear} transforms as a MW spinor and is zero for each $g\in G$.

Using this nice property of \eqref{eqn:sclinear} allows us to introduce a map between group elements $g$ and distinct solutions of the SC. In order to obtain an explicit expression for this map, we fix $\Lambda^\alpha$ to an initial value $\Lambda_0^\alpha$ and ask which different values for ${\Lambda'}^\alpha$ are possible after applying \eqref{eqn:shiftvag} to all group elements $g\in G$. The subset of elements leaving the specific choice $\Lambda_0^\alpha$ invariant are assigned to the stabilizer subgroup $H \subset G$. This permits us to decompose each group element into
\begin{equation}\label{eqn:g=mh}
  g = m h \quad \text{with} \quad h \in H
\end{equation}
and $m$ denoting a coset representative in the left coset $G/H$. The given structure allows us to identify $G$ with a $H$-principal bundle, whose tangent space fits into the exact sequence
\begin{equation}\label{eqn:seqcoset}
\begin{tikzpicture}
  \matrix (m) [matrix of math nodes, column sep=3em, row sep=1.5em] {%
    0 & \mathfrak{h} & T G & T G/H & 0 \,. \\
      &              & G   & G/H   & \\};
  \draw[->] (m-1-1) -- (m-1-2.west |- m-1-1);
  \draw[->,shiftup] (m-1-2.east) -- (m-1-3.west |- m-1-2) node[midway, above] {$\sharp$};
  \draw[<-,shiftdn] (m-1-2.east) -- (m-1-3.west |- m-1-2) node[midway, below] {$\omega$};
  \draw[->,shiftup] (m-1-3.east) -- (m-1-4.west |- m-1-3) node[midway, above] {$\pi_*$};
  \draw[<-,shiftdn] (m-1-3.east) -- (m-1-4.west |- m-1-3) node[midway, below] {${\sigma_i}_*$};
  \draw[->] (m-1-4) -- (m-1-5.west |- m-1-4);
  \draw[->,shiftup] (m-2-3.east) -- (m-2-4.west |- m-2-3) node[midway, above] {$\pi$};
  \draw[<-,shiftdn] (m-2-3.east) -- (m-2-4.west |- m-2-3) node[midway, below] {$\sigma_i$};
  \draw[->] (m-1-3.south) -- (m-2-3.north);
  \draw[->] (m-1-4.south) -- (m-2-4.north);
\end{tikzpicture}
\end{equation}
In the following, we fix all maps which appear in this diagram. Each element in $G$ is parameterized by the coordinates $X^I$. Furthermore, we assign to each coset representative $m$ the coordinates $x^i$ and to an element $h\in H$ of the subgroup $\tilde x^{\tilde i}$. By doing so, we explicitly implement the splitting
\begin{equation}
  X^I = \begin{pmatrix} x^i & \tilde x^{\tilde i} \end{pmatrix}
    \quad \text{with} \quad
  I = 1,\,\dots,\,\mathrm{dim}\,G\,, \quad
  i = 1,\,\dots,\,\mathrm{dim}\,G/H \quad \text{and} \quad
  \tilde i = 1,\,\dots,\,\mathrm{dim}\, H
\end{equation}
of the coordinates on the group manifold. In terms of these adapted coordinates the projection
\begin{equation}
  \pi(X^I) = x^i
\end{equation}
only chops the $\tilde x^{\tilde i}$ part of $X^I$. From this equation, we deduce the corresponding differential map
\begin{equation}
  \pi_*( V^I \partial_I ) = V^i \partial_i\,.
\end{equation}
The tangent space $T_g G$ at an arbitrary point of the group manifold is linked to a Lie algebra element by the left-invariant Maurer Cartan form
\begin{equation}\label{eqn:linvmcform}
  (\omega_L)_g = g^{-1} \partial_I g \, d X^I = t_A E^A{}_I d X^I
\end{equation}
which is identical with the background generalized vielbein. Moreover, there exists an isomorphism $\sharp$ between left invariant vector fields and Lie algebra elements. It has the property $\omega_L ( t_A^\sharp ) = t_A$ and reads
\begin{equation}
  t_A^\sharp = E_A{}^I \partial_I\,.
\end{equation}
This fixes all maps in \eqref{eqn:seqcoset} going from left to right. The opposite way is more involved.

In order to find the remaining maps $\sigma_i$ and $\omega$, first note that generators in the Lie algebra $\mathfrak{g}=\mathfrak{m}\oplus\mathfrak{h}$ can be split into two complement sets
\begin{equation}
  t_A = \begin{pmatrix} t_a & t^a \end{pmatrix}
    \quad \text{with} \quad
  t_a \in \mathfrak{m}
    \quad \text{and} \quad
  t^a \in \mathfrak{h}
\end{equation}
if $H$ is a maximal isotropic subgroup. Employing this splitting to the left-invariant Maurer-Cartan form \eqref{eqn:linvmcform} with the group element \eqref{eqn:g=mh}, we find
\begin{equation}\label{eqn:compbgvielbein}
  E^A{}_I = \begin{pmatrix} E_a{}_i & 0 \\
    E^a{}_i & E^a{}_{\tilde i}
  \end{pmatrix} \quad \text{and} \quad
  E_A{}^I = \begin{pmatrix} 0 & E^a{}^{\tilde i} \\
    E_a{}^i & E_a{}^{\tilde i}
  \end{pmatrix}\,.
\end{equation}
A quick calculation confirms the property $\pi_* ( {t^a}^\sharp ) = E^{ai} \partial_i = 0$ of the exact sequence \eqref{eqn:seqcoset}. Next, we take a closer look at the $\mathfrak{h}$-valued connection one-form $\omega$ of the $H$-principal bundle. It has a nice geometric interpretation as splitting the tangent space $T G = H G \oplus V G$ into a horizontal part $H G$ and a vertical one $V G$. While the latter is spanned by the fundamental vector field $(t^a)^\sharp$, the former is defined as the kernel
\begin{equation}
  H G = \{ X \in T G \,|\, \omega(X) = 0 \}
\end{equation}
of the connection. For this splitting to be consistent, the connection has to fulfill the two constraints
\begin{equation}\label{eqn:constrconnection}
\omega( {t^a}^\sharp ) = t^a \quad \text{and} \quad
  R_h^* \omega = Ad_{h^{-1}} \omega
\end{equation}
where $R_g$ denotes right translations $R_h g = g h$ on the group manifold. We want to fix the connection in such a way that the resulting subspace $H G$ of $T G$ solves the linear constraint \eqref{eqn:sclinear}. To this end, we define a projector $P_m$ at each point $m$ of the coset space $G/H$. It is a map
\begin{equation}
  P_m: \mathfrak{g} \rightarrow \mathfrak{h}, \quad P_m = t^a (P_m)_{aB} \theta^B\,,
\end{equation}
with the important property
\begin{equation}\label{eqn:Pmta}
  P_m t^a = t^a \quad \forall t^a \in \mathfrak{h}\,.
\end{equation}
Remember, $\theta^A$ is the one-form dual to the Lie algebra generator $t_A$ define in \eqref{eqn:gen&dual}. In order to fix a distinct solution of the SC on the complete group manifold, we extrapolate this projector from the coset $G/H$ to $G$ by the prescription
\begin{equation}\label{eqn:Pg}
  P_g = P_{m h} = \mathrm{Ad}_{h^{-1}} P_m \mathrm{Ad}_h\,.
\end{equation}
Now we have everything we need to finally define a connection one-form
\begin{equation}
  \omega_g = P_g \, {\omega_L}_g
\end{equation}
with the desired properties \eqref{eqn:constrconnection}.

Last but not least we come to the local sections $\sigma_i$, which are defined on the patches $U_i \subset G/H$ covering the coset space. The most general local section
\begin{equation}
  \sigma_i (x^j) = \begin{pmatrix} \delta^j_k x^k & f_i^{\tilde j} \end{pmatrix}
\end{equation}
is specified by a set of $\tilde j=1,\dots,\mathrm{dim}\,H$ functions $f^{\tilde j}_i$ in each patch. It fulfills $\pi \sigma_i = \mathrm{id}_{G/H}$ and its differential map reads
\begin{equation}
  {\sigma_i}_* (v^j \partial_j) = v^j \partial_j + v^k \partial_k f_i^{\tilde j} \tilde\partial_{\tilde j} \,.
\end{equation}
We are now able to calculate the canonical pullback of the connection one-form $A_i^0 = {\sigma_i^0}^* \omega$ with $f_i^{\tilde j} = 0$ in every patch $U_i$. For more general local sections with arbitrary functions $f_i^{\tilde j} \ne 0$, we obtain
\begin{equation}
  A_i = h_i^{-1} d h_i + h_i^{-1} A^0_i h_i
    \quad \text{with} \quad
  h_i(x^j) = h\big(f_i^{\tilde j}(x^k)\big) \,.
\end{equation}
This a gauge transformation for the gauge potential $A_i$. If we can write $A^0_i$ as
\begin{equation}
  A^0_i = h_i d h_i^{-1}
\end{equation}
there is a particular choice of the functions $f_i^{\tilde j}$ such that the pullback $A_i$ vanishes. An equivalent statement is that the field strength
\begin{equation}\label{eqn:F=0}
  F^0_i (X,Y) = d A^0_i (X,Y) + [A^0_i(X), A^0_i(Y)] = 0
\end{equation}
vanishes. If this happens in all patches, we found a flat connection. This is exactly the case we are interested in. Now, the sequence \eqref{eqn:seqcoset} becomes exact also from right to left because $\omega {\sigma_i}_* = 0$. In the overlap region $U_i \cap U_j$ between two patches the local sections are connected by
\begin{equation}
  \sigma_j = \sigma_i t_{ij} 
\end{equation}
where $t_{ij}: U_i \cap U_j \rightarrow H$ denotes the transition functions of the bundle. Because the connection one-form is uniquely defined on G, we obtain the compatibility condition
\begin{equation}
  A_j = t_{ij}^{-1} A_i t_{ij} + t_{ij}^{-1} d t_{ij}\,.
\end{equation}

The condition $\omega {\sigma_i}_* = 0$ only fixes the section locally. In general there is no global section (all transition functions $t_{ij}=e$ are the identity $e\in H$). A principal $H$-bundle is trivial iff it admits a global section. Otherwise, we are not able to split $T G = V G \oplus H G$ globally. Still, non-trivial $H$-bundles are not an obstruction in solving the SC globally. In order to prove this statement, assume that we already solved the SC locally in all patches. This is equivalent to find coordinates on each patch $U_i$ in which the curved version of $\eta^{IJ}$ has the form
\begin{equation}\label{eqn:formSCS}
  \eta^{IJ} = \begin{pmatrix} 0 & \bullet \\ \bullet & \bullet \end{pmatrix}\,,
    \quad \text{implying} \quad
  D_A f(x) D^A g(x) = \eta^{IJ} \partial_I f(x) \partial_J g(x) = 0\,.
\end{equation}
Here $\bullet$ is a placeholder for in general non-vanishing contributions. Coordinates in overlapping patches $U_i$ and $U_j$ are connected to each other by the transition function $\varphi_{ij}: U_i \rightarrow U_j$. On $\eta^{IJ}$ it acts as
\begin{equation}
  \eta^{IJ} \rightarrow \partial_M \varphi^I_{ij} \eta^{MN} \partial_N \varphi^J_{ij}\,.
\end{equation}
As long as $\varphi_{ij}$ is restricted to
\begin{equation}\label{eqn:varphiij}
  \varphi_{12}^i = \varphi_{12}^i (x) \quad \text{and} \quad
  \varphi_{12}^{\tilde i} = \varphi_{12}^{\tilde i} (x, \tilde x)\,,
\end{equation}
the form of $\eta^{IJ}$ in \eqref{eqn:formSCS} is preserved and we obtain a global SC solution. Because $H$ acts freely on $G$, the coset space $M=G/H$ is a differentiable manifold and its coordinates $x^i$ automatically satisfy the first constraint in \eqref{eqn:varphiij}. For $\varphi_{12}^{\tilde i}$, there is no restriction. Therefore, patching by arbitrary gauge transformations does not obstruct a globally defined SC solution. Later arises if, and only if, the $H$-principal bundle is flat.

Note that the connection $F$ on this bundle was introduced as a tool to discuss solutions of the SC. There is more familiar connection on the $H$-principal bundle. It results from restricting the left-invariant Maurer-Cartan form to the subalgebra $\mathfrak{h}$. Pulling it back to $T^* G/H$, we obtain the gauge potential
\begin{equation}\label{eqn:Acal}
  \mathcal{A}_i = t^a (E_i)_{a j}\, d x^j\
\end{equation}
where $(E_i)_A{}^J$ denotes the background generalized vielbein in the patch $i$. We will see while presenting examples in section~\ref{sec:examples} that its field strength $\mathcal{F}_i = d \mathcal{A}_i + [\mathcal{A}_i, \mathcal{A}_i]$ in general does not vanish. Even if the SC is solved globally. It can be used to classify the bundle in terms of its characteristic classes.

\subsection{Explicit Form of the Connection}\label{sec:linkspinoromega}
In this subsection we take a closer look at the projector $P_m$ which is used to fix the connection one-form $\omega$ on $G$. So far, we treated it as an abstract object by only requiring its elementary property \eqref{eqn:Pmta}. Now, we construct it explicitly. 

To this end, we first relate the $\Gamma$-matrices in \eqref{eqn:sclinear} to fermionic creation/annihilation operators $\psi^a$/$\psi_a$. They fulfill the canonical anti-commutator relations
\begin{equation}
  \{\psi_a, \psi^b\} = \delta_a^b \quad \text \quad
  \{\psi_a, \psi_b\} = \{\psi^a, \psi^b\} = 0\,.
\end{equation}
If we scale them accordingly, they are equal to
\begin{equation}
  \Gamma_a = \sqrt{2} \psi_a \quad \text{and} \quad
  \Gamma^a = \sqrt{2} \psi^a
\end{equation}
and reproduce the Clifford algebra \eqref{eqn:cliffordalg}. We define the pure spinor $|\Lambda_0\rangle$ from the last subsection to be the normalized vacuum state of this setup
\begin{equation}
  \langle \Lambda_0 | \Lambda_0 \rangle = 1\,.
\end{equation}
Thus, it is annihilated by
\begin{equation}
  \psi_a |\Lambda_0\rangle = 0\,.
\end{equation}
In the following we suppress spinor indices and use a bra-ket notation instead. Hermitian conjugation relates creation and annihilation operators
\begin{equation}
  ( \psi^a )^\dagger = \psi_a\,.
\end{equation}
That is all we need to write down the first version of
\begin{equation}\label{eqn:Pmone}
  P_m = \frac{t_A}2 \langle \Lambda_0 | \Gamma^A \Gamma_B | \Lambda_0 \rangle \theta^B = t^a \delta_a^b \theta_b = t^a \theta_a\,.
\end{equation}
As one immediately sees, it is a trivial implementation of the property \eqref{eqn:Pmta}.

We can do better. A MW spinor of Pin($d,d$), the double cover of O($d,d$), has $2^{d-1}$ real components. There is an isomorphism between them
\begin{equation}\label{eqn:expansionMW}
  |\Lambda\rangle = \sum\limits_{n=0}^{\lfloor d/2 \rfloor} \frac{1}{2^n (2 n) !} B^{(2n)}_{a_1 \dots a_{2n}} \Gamma^{a_1 \dots a_{2n}} | \Lambda_0 \rangle
\end{equation}
and the even, totally anti-symmetric $2n$ forms $B^{(2n)}$ up to degree $2 \lfloor d/2 \rfloor$. Replacing the right-hand side of \eqref{eqn:Pmone} with this spinor results in
\begin{equation}\label{eqn:Pmtwo}
  P_m = \frac{t_A}2 \langle \Lambda_0 | \Gamma^A \Gamma_B | \Lambda \rangle \theta^B = \frac{t^A}2 \langle \Lambda_0 | \Gamma^A \Gamma_B \Big( 1 + \frac{1}{4} B_{c d} \Gamma^{cd} \Big) | \Lambda_0 \rangle \theta^B \,.
\end{equation}
Let us explain this expression. First, the constraint \eqref{eqn:Pmta} fixes the $B^{(0)}$ contribution to be one. At most there are two annihilation operators on the left-hand side of the bra-ket. They can only compensate two creation operators on the right. Thus, all contributions $B^{(2n)}$ from \eqref{eqn:expansionMW} with $n>1$ do not contribute to \eqref{eqn:Pmtwo}. This leaves us with $B^{(2)}$ which we rename to $B$. In order to simplify this result further, one replaces the $\Gamma$-matrices by the corresponding creation/annihilation operators and swaps them until all creation operators are on the left. This procedure results in the simple expression
\begin{equation}
  P_m = - t^a B_{ab} \theta^b + t^a \theta_a\,.
\end{equation}
By construction, this projector has all SC solutions for $\Lambda$ as its nullspace. After combining it with the left-invariant Maurer-Cartan form, we obtain the gauge potential
\begin{equation}
  A = t^a ( - B_{ab} E^b{}_i + \delta_a^b E_{bi} )\, d x^i\,.
\end{equation}
Note that we suppress the index labeling the patch dependence of the section for the sake of brevity here and also in the following. For $A=0$, we are able to reconstruct the solution of the SC on each point of the coset $G/H$ in terms of the two-form
\begin{equation}\label{eqn:BfromE}
  B = - \frac12 \delta^b_a E_{bi} E^a{}_j \, d x^i\wedge d x^j\,.
\end{equation}

This result is interesting, because it provides a shortcut to obtain a vanishing connection. Normally, we would first try to fix the two-form $B$ in such a way that the field strength $F$ in \eqref{eqn:F=0} vanishes. In this case $A$ is a pure gauge and one can set it to zero by an appropriate gauge transformation in every patch. An equivalent statement is that the symmetric part of $\delta^b_a E_{bi} E^a{}_j$ vanishes. This is the case iff
\begin{equation}
  2 E_{b(i} \delta^b_a E^a{}_{j)} = E^A{}_i \eta_{AB} E^B{}_j = \eta_{ij} = 0
\end{equation}
holds. In the following, we show that $\eta_{ij}=0$ automatically holds if: First, we choose a coset representative of the form
\begin{equation}\label{eqn:formsol}
  m = \exp [ f ( x^i) ] \quad \text{with} \quad f: U \rightarrow \mathfrak{m}
\end{equation}
where $U$ denotes a patch of the physical subspace $G/H$. And second, the two complement sets $\mathfrak{g}=\mathfrak{m}\oplus\mathfrak{h}$ spanned by the generators $t_a$ and $t^a$ fulfill the relations
\begin{equation}\label{eqn:symmetricpair}
  [\mathfrak{h},\mathfrak{h}] \subset \mathfrak{h} \,, \quad
  [\mathfrak{h},\mathfrak{m}] \subset \mathfrak{m} \quad \text{and} \quad
  [\mathfrak{m},\mathfrak{m}] \subset \mathfrak{h}\,.
\end{equation}
They render $\mathfrak{m}$ and $\mathfrak{h}$ to be a symmetric pair and consequently the coset $G/H$ is a symmetric space. In order to keep the following expressions compact, it is convenient to introduce the bilinear inner product
\begin{equation}\label{eqn:bilinearform}
  \langle t_A, t_B \rangle = \eta_{AB}\,.
\end{equation}
In this notation, we have to show that
\begin{equation}\label{eqn:etaijformula}
  \eta_{ij} = \langle m^{-1} \partial_i m , m^{-1} \partial_j m  \rangle = 0\,.
\end{equation}
In order to evaluate this expression, we use the Baker-Campbell-Hausdorff relation
\begin{equation}
  e^f t e^{-f} = \sum_{n=0}^\infty \frac{1}{n!} [f,t]_n
    \quad \text{with} \quad
  [f,t]_n = [\underbrace{f [\dots, [f}_{n\text{ times}},t] \dots ]]\,,
\end{equation}
which gives rise to
\begin{equation}
  m \partial_i m^{-1} = \sum_{n=0}^\infty \frac{1}{(n+1)!} [f,\partial_i f]_n\,.
\end{equation}
Furthermore, we exploit that $\mathfrak{g}$ can be embedded into $\mathfrak{o}$($d,d$) to identify
\begin{equation}
  \langle t_A, [t_B, t_C] \rangle = \langle [t_A, t_B], t_C \rangle
\end{equation}
which after several iterations allows to derive the relation
\begin{equation}
  \langle [t_A, t_B]_m, [t_A, t_C]_n \rangle = - \langle( [t_A, t_B]_{m+1}, [t_A, t_C]_{n-1} \rangle\,.
\end{equation}
After applying it $n-m$ times, where $n>m$, we obtain
\begin{equation}
  \langle [t_A, t_B]_m, [t_A, t_C]_n \rangle = (-1)^{n+m} \langle [t_A, t_B]_n, [t_A, t_C]_m \rangle\,.
\end{equation}
Now we are able to decompose the expansion for \eqref{eqn:etaijformula} into the two contributions
\begin{align}
  \eta_{ij} = \langle m \partial_i m^{-1}, & m \partial_j m^{-1} \rangle = \sum_{m=0}^\infty \frac{1}{\big((m+1)!\big)^2} \langle [f, \partial_i f]_m, [f, \partial_j f]_m \rangle + \nonumber \\
  &\quad \sum_{m=1}^\infty \sum_{n=0}^{m-1} \frac{1 + (-1)^{m+n}}{(m+1)! (n+1)!} \langle [f, \partial_i f]_m, [f, \partial_j f]_n \rangle\,.
\end{align}
Keeping in mind \eqref{eqn:symmetricpair} and $\langle t_a, t_b \rangle = \langle t^c, t^d \rangle = 0$, it is straightforward to show that
\begin{equation}\label{eqn:vanishingK}
  \langle [f,\partial_i f]_m, [f,\partial_j f]_n \rangle = 0 \quad \text{for } n + m \text{ is even.}
\end{equation}
Finally, we find the desired result
\begin{equation}
  \eta_{ij} = 0\,.
\end{equation}
Hence, we are able to construct a flat connection for all group manifolds whose Lie algebras admit a splitting into a symmetric pair. Additionally the subalgebra $\mathfrak{h}$ of this pair has to be maximal isotropic or equivalently there exists an initial solution of the SC $\Lambda_0$.

\subsection{Generalized Geometry}\label{sec:gg}
Looking closely at the exact sequence \eqref{eqn:seqcoset}, we find a striking resemblance with an exact Courant algebroid based on the exact sequence
\begin{equation}\label{eqn:exactCourantAlg1}
\begin{tikzpicture}
  \matrix (m) [matrix of math nodes, column sep=3em, row sep=1.5em] {%
    0 & T^* M & E & T M & 0 \\};
  \draw[->] (m-1-1) -- (m-1-2.west |- m-1-1);
  \draw[->] (m-1-2.east) -- (m-1-3.west |- m-1-2);
  \draw[->] (m-1-3.east) -- (m-1-4.west |- m-1-3);
  \draw[->] (m-1-4) -- (m-1-5.west |- m-1-4);
\end{tikzpicture}
\end{equation}
after identifying $M=G/H$ and $E = T G$. In order to obtain a complete match we still need an isomorphism between $\mathfrak{h}\cong V G$ and $T^* M$. It is mediated by the bijective map $\eta: \mathfrak{h} \rightarrow T^* M$ which is defined as
\begin{equation}
  \eta = \theta_a E^a{}_i \, d x^i\,.
\end{equation}
Its inverse $\eta^{-1}: T^*M \rightarrow \mathfrak{h}$ reads
\begin{equation}
  \eta^{-1} = t^a i_{E_a}
\end{equation}
where $E_a$ denotes the vector $E_a{}^i \partial_i$. In conjugation with the maps we have discussed in subsection \ref{sec:HPrincipal}, $\eta$ allows us to refine the exact sequence \eqref{eqn:exactCourantAlg1} to \eqref{eqn:exactCourantAlg2} in the introduction. All maps appearing there are nothing else than the components of a generalized frame field $\hat E_A$ and its inverse $\hat E^A$ \cite{Hohm:2010xe,Grana:2008yw,Lee:2014mla}. It represents a map from $E\cong\mathfrak{g}$ to the generalized tangent bundle $T^* M \oplus T M$\footnote{As we pointed out in the introduction, this bundle can be twisted by a gerbe introduced through the $B$-field in $\hat E_A$. In this case one should maintain the name $E$ for generalized tangent bundle instead of $T^* M \oplus T M$}. We immediately read off its components
\begin{equation}
\label{eqn:vielbeinhat1}
  \hat E_A{} = \left. \Big( \pi_* (t_A^\sharp) + \eta \omega( t_A^\sharp) \Big)\right|_{\sigma(x^i)} \,.
\end{equation}
The corresponding dual frame follows directly from the properties of the exact sequence and reads
\begin{equation}
  \hat E^A{} (v, \tilde v) = \left. \theta^A \Big( \eta^{-1}(\tilde v) + i_{\sigma_* v} \omega_L \Big) \right|_{\sigma(x^i)}\,.
\end{equation}
Here, we denote elements of the generalized tangent bundle as $V = v + \tilde v$ with $v \in T M$ and $\tilde v \in T^* M$. Finally, let us write down the generalized frame and its dual in the explicit form
\begin{equation}\label{eqn:genframe&dual}
  \hat E_A = \begin{pmatrix}
    E^a{}_i \, d x^i & & \\
    B_{ab} E^b{}_i \, d x^i & + & E_a{}^i \partial_i \end{pmatrix}
  \quad \text{and} \quad
  \hat E^A (v,\tilde v) = \begin{pmatrix}
    E_a{}^i \tilde v_i & + & B_{ab} E^b{}_i v^i \\
    & & E^a{}_i d x^i
    \end{pmatrix}\,.
\end{equation}
In order to obtain the dual frame, one has to take into account $\sigma^* \omega = 0$ which results in
\begin{equation}
  \theta_a \omega_L ( \sigma_* v ) = B_{ab} E^b{}_i v^i\,.
\end{equation}
This result makes perfect sense, because it reproduces the canonical vielbein of DFT \cite{Hohm:2010xe,Geissbuhler:2011mx,Geissbuhler:2013uka}
\begin{equation}
  \hat E_A{}^{\hat I} = \begin{pmatrix} E^a{}_i & 0 \\
    E_a{}^i B_{ij} & E_a{}^i
  \end{pmatrix} \quad \text{and its inverse transposed} \quad
  \hat E^A{}_{\hat I} = \begin{pmatrix} E_a{}^i & E_a{}^i B_{ij} \\
    0 & E^a{}_i
  \end{pmatrix}\,. 
\end{equation}
At this point it is obvious that the curved version $B_{ij}= B_{ab} E^a{}_i E^b{}_j$ of the two-form capturing the solution of the SC on $G$ is nothing else than a $B$-field in DFT. Additionally, it is convenient to label elements of the generalized tangent bundle with hatted indices $\hat I, \hat J, \hat K, \dots$\,. These are defined as
\begin{equation}
  V^{\hat I} = \begin{pmatrix} \tilde v_i & v^i \end{pmatrix}
    \quad \text{and} \quad
  V_{\hat I} = \begin{pmatrix} v^i & \tilde v_i \end{pmatrix}
    \quad \text{with} \quad
  i=1,\,\dots,\,d
\end{equation}
in complete analogy with the standard formulation of DFT. Thus, applying the dual frame field $\hat E^A$ to $\eta_{AB}$, we obtain
\begin{equation}
  \eta_{\hat I\hat J} = \hat E^A{}_{\hat I} \eta_{AB} \hat E^B{}_{\hat J} = \begin{pmatrix} 0 & \delta_j^i \\
    \delta_i^j & 0 \end{pmatrix}
\end{equation}
which also shows that $\hat E_A{}^{\hat I}$ and $\hat E^A{}_{\hat I}$ are O($d,d$) elements.

With the generalized frame and its inverse, we are able to pull the generalized Lie derivative of \DFTwzw{} \eqref{eqn:genLie} to the generalized tangent bundle. Its elements are connected to $V^A$ by the relation
\begin{equation}
  V^{\hat I} = V^A \hat E_A{}^{\hat I}\,.
\end{equation}
Applying the dual frame on the flat derivative $D_A$, we further obtain
\begin{equation}
  \partial_{\hat I} = \hat E^A{}_{\hat I} D_A = \begin{pmatrix} \partial_i & 0 \end{pmatrix}\,.
\end{equation}
Plugging these two relations in the generalized Lie derivative \eqref{eqn:genLie} written in terms of flat derivatives
\begin{equation}
  \mathcal{L}_\xi V^A = \xi^B D_B V^A + ( D^A \xi_B - D_B \xi^A ) V^B + F_{BC}{}^A \xi^B V^C
\end{equation}
gives rise to
\begin{equation}\label{eqn:genlie0}
  \mathcal{L}_\xi V^{\hat I} = \mathcal{L}^0_\xi V^{\hat I} + \mathcal{F}_{\hat J\hat K}{}^{\hat I} \xi^{\hat J} V^{\hat K}
\end{equation}
where
\begin{equation}
  \mathcal{L}^0_\xi V^{\hat I} = \xi^{\hat J} \partial_{\hat J} V^{\hat I} + ( \partial^{\hat I} \xi_{\hat J} - \partial_{\hat J} \xi^{\hat I} ) V^{\hat J}
\end{equation}
denotes the standard untwisted generalized Lie derivative of DFT and $\mathcal{F}_{\hat I\hat J}{}^{\hat K}$ is a twist. It is the curved version $\mathcal{F}_{\hat I\hat J}{}^{\hat K} = \mathcal{F}_{AB}{}^C \hat E^A{}_{\hat I} \hat E^B{}_{\hat J} \hat E_C{}^{\hat K}$ of
\begin{equation}\label{eqn:scFABC}
  \mathcal{F}_{AB}{}^C = F_{AB}{}^C - \mathcal{L}^0_{\hat E_A} \hat E_B{}^{\hat I} \hat E^C{}_{\hat I}
\end{equation}
and combines the structure coefficients of the Lie algebra $\mathfrak{g}$ with the covariant fluxes of the generalized frame field $\hat E_A$. Pulled to the generalized tangent bundle, the structure coefficients $F_{\hat I\hat J\hat K}$ play the role of fluxes. We use the following conventions 
\begin{align}
  H &= \frac1{3!} F_{ijk} \, d x^i \wedge d x^j \wedge d x^k &
  f &= \frac1{2!} F_{ij}{}^k \, d x^i \wedge d x^j \wedge \partial_k \nonumber \\
  Q &= \frac1{2!} F^{ij}{}_k \, \partial_i \wedge \partial_j \wedge d x^k  &
  R &= \frac1{3!} F^{ijk} \, \partial_i \wedge \partial_j \wedge \partial_k
\end{align}
to write down the components of the $H$-, $f$-, $Q$- and $R$-flux in a compact form. We apply the same notation to the twist, for which the fluxes are decorated by the subscript $\mathcal{F}$, and to the contributions from the generalized frame field labeled by the subscript $\hat E$.

\subsection{Deformations}\label{sec:deformations}
In the last subsections, we constructed solutions of the SC and linked them to GG. But can the formalism presented in this paper provide any additional information which are not manifest in a GG description? One way to approach this question, is to check whether the SC solution for a given group manifold $G$ is unique. If it is not, there exist different GGs associated to the same doubled space. That is what DFT is supposed to do. It connects different target spaces which are dual to each other. In the standard formulation, different solutions of the SC are in one-to-one correspondence with a subset of $O(d,d)$ transformations. For example, certain O($d$)$\times$O($d$)$\subset$O($d,d$) transformations applied to the DFT generalized metric reproduce the Buscher rules \cite{Buscher:1987sk}. Here the situation is more general. We do not have to talk the  generalized metric on the doubled space into account. We only fix its topology by specifying a group manifold $G$. Different solutions of the SC are captured by deformations of the maximal isotropic subalgebra $\mathfrak{h}\subset\mathfrak{g}$ and describe, as we show explicitly in section~\ref{sec:examples}, dual target spaces with different topologies. They are conveniently classified in terms of Lie algebra cohomology, reviewed in appendix~\ref{app:liealgebracohomology}. 

One can freely choose the MW spinor $\Lambda^0$ as long as the generators $t^a$ which violate the SC form a subalgebra $\mathfrak{h}\subset\mathfrak{g}$. An arbitrary MW spinor can be brought to a canonical form by an O($d,d$) transformation. Thus instead of changing $\Lambda^0$, we can also deform the generators or equivalently the structure coefficients $F_{AB}{}^C$ of the Lie algebra $\mathfrak{g}$. Doing so, we of course have to check whether $\mathfrak{h}$ still is a subalgebra. Let us take a closer look on how this works. First of all, we only have to consider transformations in the coset O($d$)$\times$O($d$)/O($d$). All others would not change $\Lambda^0$ at all or only scale it. But under these operations the subalgebra $\mathfrak{h}$ is invariant. At the same time, we are able to apply arbitrary O($d,d$)/O($d$)$\times$O($d$) transformations without changing the solution of the SC. Equivalent one is completely free in  choosing a generalized metric on $G$.

Assume that we have a coset element 
\begin{equation}\label{eqn:deformrepr}
  \mathcal{T}_A{}^B = \exp ( \lambda t_A{}^B )\,,
\end{equation}
generated by applying the exponential map to a $\mathfrak{o}(d)$$\times$$\mathfrak{o}(d)$ generator $t_A{}^B$. It modifies the structure coefficients of the Lie algebra according to
\begin{equation}\label{eqn:deformationFABC}
  F'_{AB}{}^C = \mathcal{T}_A{}^D \mathcal{T}_B{}^E F_{DE}{}^F \mathcal{T}_F{}^C
\end{equation}
which gives rise to the expansion
\begin{equation}\label{eqn:defseries}
  F'_{AB}{}^C = F_{AB}{}^C + \lambda \delta F_{AB}{}^C + \lambda^2 \delta^2 F_{AB}{}^C + \dots\,.
\end{equation}
From this series, we immediately read off the $\mathfrak{g}$-valued two-forms 
\begin{equation}
  c_n = t_C ( \delta^n F_{AB}{}^C ) \theta^A \wedge \theta^B
\end{equation}
by comparing with \eqref{eqn:deformedbracket}. We are only interested in those of them which do not spoil the subalgebra $\mathfrak{h}$. Thus only transformations with $\delta^n F^{abc} = 0$ are allowed. Finally, we have to check whether the restricted forms
\begin{equation}
  c_n = t^c ( \delta^n F^{ab}_c ) \theta_a \wedge \theta_b 
\end{equation}
fulfill the constraints for a valid deformation of $\mathfrak{h}$ outlined in appendix~\ref{app:liealgebracohomology}.

\section{Examples}\label{sec:examples}
In this section, we present explicit examples for the techniques discussed in the last subsection. We stick to $d=3$, where all Lie algebras $\mathfrak{g}$ which admit an embedding into $\mathfrak{o}(3,3)$ are classified \cite{Dibitetto:2012rk} and the corresponding group manifolds $G$ are constructed \cite{Bosque:2015jda}. Our first example is the canonical one, the torus with $H$-flux. It is generated by the Lie algebra $\mathfrak{cso}(1,0,3)$ and gives rise to a chain of T-dual backgrounds which were studied extensively in the literature \cite{Shelton:2005cf,Dabholkar:2005ve,Hull:2006va,Hull:2006qs,Shelton:2006fd}. Except for the torus with $R$-flux, all of them are captured by deformations of the maximal isotropic subalgebra $\mathfrak{h}$ as described in section~\ref{sec:deformations}. Furthermore, we consider the standard example of \DFTwzw{}, the three-sphere $S^3$ with $H$-flux\footnote{See \cite{Plauschinn:2013wta,Plauschinn:2014nha} for a recent discussion of T-duality in this background.}. Here, the Lie algebra is $\mathfrak{so}(4)$ and its subalgebra $\mathfrak{h}=\mathfrak{so}(3)$ does not admit non-trivial deformations. However, it is still possible to find a T-dual solution of the SC after removing a discrete subgroup from the physical space $M$=SO($4$)$/$SO($3$). All results are in perfect agreement with the topology changes one would expect from T-duality \cite{Bouwknegt:2003vb}. By calculating the characteristic classes of the $H$-principle bundles of these examples, we also see that the generalized tangent bundle is in general twisted.

\subsection{Torus with \texorpdfstring{$H$}{H}-Flux}\label{sec:torusHflux}
A torus with $\mathbf{h}$ units of $H$-flux is characterized by the Lie algebra $\mathfrak{g}$=$\mathfrak{cso}(1,0,3)$ with the non-vanishing commutator relations\footnote{We use the convention $\epsilon_{123} = 1$ for the totally antisymmetric tensor. Its indices are raised and lowered with the identity $\delta^{ab}$/$\delta_{ab}$.}
\begin{equation}\label{eqn:LiealgHtorus}
  [t_a, t_b] = \mathbf{h} \, \epsilon_{abc} t^c \,.
\end{equation}
The three generators $t^1$, $t^2$, $t^3$ are in the center of this algebra and form an abelian, maximal isotropic subalgebra $\mathfrak{h}$. In order to construct a SC solution for this background, we first obtain all elements of $G$ by applying the exponential map to a faithful matrix representation of the Lie algebra \cite{Bosque:2015jda}, resulting in
\begin{equation}\label{eqn:paramCSO(1,0,3)}
  m = \exp( t_1 x^1 ) \exp( t_2 x^2 ) \exp( t_3 x^3 ) \quad \text{and} \quad
  h = \exp( t^1 \tilde x^1 ) \exp( t^2 \tilde x^2 ) \exp( t^3 \tilde x^2 )\,.
\end{equation}
Note that the group CSO($1,0,3$) we get from this procedure is not compact. In order to compactify it, we mod out the discrete subgroup CSO($1,0,3,\mathbb{Z}$) (acting from the left) which results in the identifications
\begin{align}
  (x^1\,,x^2\,,x^3\,,\tilde x^1\,,\tilde x^2\,, \tilde x^3) &\sim (x^1 + 1\,, x^2\,, x^3\,, \tilde x^1\,, \tilde x^2\,, \tilde x^3) \nonumber \\
  & \sim (x^1\,, x^2 + 1\,, x^3\,, \tilde x^1\,, \tilde x^2\,, \tilde x^3 - x^1 \mathbf{h}) \nonumber \\
  & \sim (x^1 \,, x^2\,, x^3 + 1\,, \tilde x^1 - x^2 \mathbf{h}\,, \tilde x^2 + x^1 \mathbf{h}\,, \tilde x^3) \nonumber \\
  & \sim (x^1\,, x^2\,, x^3\,, \tilde x^1 + 1\,, \tilde x^2\,, \tilde x^3) \nonumber \\
  & \sim (x^1\,, x^2\,, x^3\,, \tilde x^1\,, \tilde x^2 + 1\,, \tilde x^3) \nonumber \\
  & \sim (x^1 \,, x^2\,, x^3\,, \tilde x^1\,,\tilde x^2\,, \tilde x^3 + 1) \label{eqn:identtwithH}
\end{align}
for the coordinates on the group manifold. Their derivation is explained in appendix~\ref{sec:modoutcso(103z)}. Note that they are respecting the restriction \eqref{eqn:varphiij} on transition functions and therefore give rise to a global SC solution once the SC is solved locally in every patch. In order to find this local solution, we now follow the steps outlined in section~\ref{sec:solSC} and try to obtain a $B$-field such that the field strength $F = F_a t^a$ for the connection
\begin{align}
  A_1 &= ( \phantom{-} \mathbf{h} x^3 - B_{12} )\, d x^2 - B_{13} d x^3\,,\quad \nonumber \\
  A_2 &= ( - \mathbf{h} x^3 + B_{12} )\, d x^1 - B_{23}\, d x^3 \quad \text{and} \nonumber \\
  A_3 &= ( \phantom{-} \mathbf{h} x^2 + B_{13} )\, d x^1 + B_{23} d x^2
\end{align}
is zero. This is not very complicated because the gauge group is abelian and the field strength $F_a = d A_a$ vanishes for
\begin{equation}\label{eqn:BforTwH}
  B = \frac{\mathbf{h}}2 x^3 d x^1 \wedge d x^2 - \frac{\mathbf{h}}2 x^2 d x^1 \wedge d x^3 + \frac{\mathbf{h}}2 x^1 d x^2 \wedge d x^3\,.
\end{equation}
In this case $A_a$ is pure gauge and can be written in the form $A_a = d \lambda_a$ with
\begin{equation}
  \lambda_1 =   \frac{\mathbf{h}}2 x^2 x^3\,, \quad
  \lambda_2 = - \frac{\mathbf{h}}2 x^1 x^3 \quad \text{and} \quad
  \lambda_3 =   \frac{\mathbf{h}}2 x^1 x^2\,. 
\end{equation}
After applying the transformation $g \rightarrow g \exp (t^a \lambda_a)$ to the group elements $g = m h$, we obtain the components
\begin{equation}\label{eqn:EAiHflux}
  E_{ai} = \frac{\mathbf{h}}2 \begin{pmatrix}
    0 & \phantom{-} x^3 & - x^2 \\ - x^3 & 0 & \phantom{-} x^1 \\ \phantom{-} x^2 & - x^1 & 0
  \end{pmatrix}\,, \quad
  E^a{}_i = \begin{pmatrix}
    1 & 0 & 0 \\ 0 & 1 & 0 \\ 0 & 0 & 1
  \end{pmatrix} \quad \text{and} \quad
  E^a{}_{\tilde i} = \begin{pmatrix}
    1 & 0 & 0 \\ 0 & 1 & 0 \\ 0 & 0 & 1
  \end{pmatrix}
\end{equation}
of the background generalized vielbein defined in equation~\eqref{eqn:compbgvielbein}. In conjunction with the $B$-field, they are sufficient to completely fix the frame field $\hat E_A$. A shortcut to obtain the same result is to start from the coset representative
\begin{equation}
  m = \exp ( t_1 x^1 + t_2 x^2 + t_3 x^3 )\,.
\end{equation}
According to our discussion is section~\ref{sec:linkspinoromega} this choice has to solve the SC. Indeed, using it we obtain the same background generalized vielbein as in \eqref{eqn:EAiHflux}.

If we calculate the twist \eqref{eqn:scFABC} of the generalized Lie derivative, the only non-vanishing contribution is
\begin{equation}
  H_{\mathcal{F}} = - \frac{\mathbf{h}}2 \, d x^1 \wedge d x^2 \wedge d x^3\,.
\end{equation}
Thus, we obtain a $H$-twisted Lie derivative. This results is surprising. Why are the $\mathbf{h}$ units of $H$-flux on the torus distributed in such a strange way between the frame field and the twist?  To answer this question, we take a closer look at the contribution coming from the former. Remember equation \eqref{eqn:BfromE} for the $B$-field. We have a SC solution with $A=0$ and can apply it to obtain
\begin{equation}\label{eqn:HfromChern}
  H_{\hat E} = d B = - \frac12 \, \mathcal{F}_a \wedge E^a{}_i d x^ i\,.
\end{equation}
$\mathcal{F}_a$ is the field strength for the gauge connection $\mathcal{A}$, defined in \eqref{eqn:Acal}. Because the subgroup $\mathfrak{h}$ is abelian for this background, each of the three $\mathcal{F}_a$ represents a characteristic class of a circle bundle over $M$. As such they have to be elements of $H^2(M, \mathbb{Z})$ and indeed taking into account \eqref{eqn:EAiHflux} results in
\begin{equation}
  \mathcal{F}_1 = - \mathbf{h} \, d x^2 \wedge d x^3 \,, \quad
  \mathcal{F}_2 = \mathbf{h} \, d x^1 \wedge d x^3 \quad \text{and} \quad
  \mathcal{F}_3 = - \mathbf{h} \, d x^1 \wedge d x^2\,.
\end{equation}
Each of them gives an integer if integrated over the torus spanned by $x^2, x^3$ / $x^1, x^3$ or $x^1, x^2$, respectively. Hence, the $H$-flux contribution from the frame field has to be
\begin{equation}
  H_{\hat E} = \frac{3 \mathbf{h}}{2} \, d x^1 \wedge d x^2 \wedge d x^3\,.
\end{equation}
A non-trivial $H$-flux in $H^3(M, \mathbb{Z})$ describes a gerbe structure (see for example. \cite{Hitchin:1999fh} for an introduction). It would be interesting to better understand how the data of the $H$-principal bundle is related to this gerbe in general. From \eqref{eqn:scFABC} it is obvious that the twist and the contribution from the frame fields have to sum up to the right among of $H$-flux,
\begin{equation}\label{eqn:HtorusH}
  H = H_{\hat E} + H_{\mathcal{F}} = \mathbf{h} \, d x^1 \wedge d x^2 \wedge d x^3\,.
\end{equation}
Integrating the two contributions over $M$, we find the cohomology classes
\begin{equation}\label{eqn:classesTwithH}
  [ H_{\hat E} ] = \int_M H_{\hat E} = \frac32 \quad \text{and} \quad
  [ H_{\mathcal{F}} ] = \int_M H_{\mathcal{F}} = - \frac12\,.
\end{equation}

On way to get ride of the twist contribution $H_{\mathcal{F}}$ is to consider a partial solution of the SC. This breaks the T-duality group from O($3,3$) to the subgroup O($2,2$). For the coordinates on the group manifold $G$ which transform in the $\mathbf{6}$ of the former, this induces the branching
\begin{equation}
  \mathbf{6} \rightarrow (\mathbf{2}, \mathbf{2}) + 2 (\mathbf{1}, \mathbf{1})\,.
\end{equation}
Now remove the generator $t^1$ from the Lie algebra and only keep commutators involving the remaining ones. Doing so, we obtain a 5 dimensional Lie algebra. Its Lie group $G$ has the coordinate irreps $(\mathbf{2}, \mathbf{2}) + (\mathbf{1}, \mathbf{1})$ where the $(\mathbf{1},\mathbf{1})$ part always solves the SC. This group manifold describes the extended space (not fully doubled anymore) as doubled two-torus with coordinates $x^2$, $x^3$, $\tilde x^2$, $\tilde x^3$ fibered over a circle with coordinate $x^1$. Now, the subalgebra algebra $\mathfrak{h}$ has only two abelian generators. This removes one of the three Chern classes from equation \eqref{eqn:HfromChern} and results in the right among of $H$-flux. At the same time, the twist vanishes. As we will see in a second this modification is in agreement with the fact that we cannot perform three independent T-duality transformations and still solve the SC. Thus, all SC solutions are indeed captured by $O(2,2)$. A drawback is that one has to modify the generalized tangent space from $T^* M \oplus T M$ to an adapted version. This renders the explicit equations derived in section~\ref{sec:linkspinoromega} and \ref{sec:gg} invalid. However the general procedure outlined there still applies.

Let us now study possible deformations of this solution. To this end, we choose the coset element $\mathcal{T}_A{}^B$ generated by
\begin{equation}\label{eqn:tdeform}
  t_A{}^B = \begin{pmatrix} 0 & t^{ab} \\
    t_{ab} & 0 \end{pmatrix}  \quad \text{with} \quad
  t_{ab} = t^{ab} = \begin{pmatrix} 0 & 1 & 0 \\ -1 & 0 & 0 \\ 0 & 0 & 0 \end{pmatrix}
\end{equation}
to deform the structure coefficients of $\mathfrak{g}$ according to \eqref{eqn:deformationFABC}. For the resulting series \eqref{eqn:defseries}, we find $\delta^n F^{abc} = 0$ and only $c_n$ with even $n$ do not vanish. Thus, we substitute $\lambda \rightarrow \sqrt{\lambda}$ in \eqref{eqn:deformrepr} and $c_{2n} \rightarrow c_n$ to obtain the deformation series \eqref{eqn:deformedbracket} with the coefficients
\begin{equation}
  c_1 = 2 \mathbf{h} \, t^3 \theta_1 \wedge \theta_2\,, \quad
  c_2 = - \frac{2 \mathbf{h}}3 \, t^3 \theta_1 \wedge \theta_2\,,
  \quad \text{\dots}\,.
\end{equation}
It is non-trivial because $c_1$ is an element of $H^2(\mathfrak{h},\mathfrak{h})$. The group elements mediating this deformation are valued in SO($3$)$\times$SO($3$) with $\sqrt{\lambda}\in [0,\, 2\pi)$. In the context of T-duality transformations, we are only interested in the restriction of $\mathcal{T}_A{}^B$ to the discrete subgroup SO($3,\mathbb{Z}$)$\times$SO($3,\mathbb{Z}$). Thus, we are left with the four elements $\sqrt{\lambda} \in \frac\pi{2} \{ 0,\, 1,\, 2,\, 3\}$ generated by the transformation
\begin{align}
  t_1 & \rightarrow   t^2 & 
  t_2 & \rightarrow - t^1 & 
  t_3 & \rightarrow   t_3 & 
  t^1 & \rightarrow   t_2 & 
  t^2 & \rightarrow - t_1 & 
  t^3 & \rightarrow   t^3 \,.
\end{align}
It is the result of the concatenation of two different transformations. First, a SO($2, \mathbb{Z}$) rotation swaps $t^1$, $t^2$ and $t_1$, $t_2$, respectively. Second, two generators from $\mathfrak{m}$ are exchanged with their counterparts in $\mathfrak{h}$ according to
\begin{align}
  \mathcal{T}_{12}: &&
  t_1 & \rightarrow t^1 & 
  t_2 & \rightarrow t^2 & 
  t_3 & \rightarrow t_3 & 
  t^1 & \rightarrow t_1 &
  t^2 & \rightarrow t_2 & 
  t^3 & \rightarrow t^3 \,.
\end{align}
Only this second part is relevant to our discussion because SO($2, \mathbb{Z}$)$\subset$SO($3$) rotations do not change the subalgebra $\mathfrak{h}$. Hence, we only keep $\mathcal{T}_{12}$ which generates a T-duality transformations along the directions $x^1$, $x^2$ and results in a torus with  $Q_3^{12}$ flux. There are two other, independent choices of $t_A{}^B$ in \eqref{eqn:tdeform},
\begin{equation}
  t_{ab} = t^{ab} = \begin{pmatrix} 0 & 0 & 1 \\ 0 & 0 & 0 \\ -1 & 0 & 0 \end{pmatrix} \quad \text{and} \quad
  t_{ab} = t^{ab} = \begin{pmatrix} 0 & 0 & 0 \\ 0 & 0 & 1 \\ 0 & -1 & 0 \end{pmatrix}\,.
\end{equation}
They mediate T-dualities along the directions $x^1$, $x^3$ and $x^2$, $x^3$, respectively. After applying any of them, $\mathfrak{h}$ is equivalent to the Heisenberg algebra. Starting over again from this algebra, we do not find any other non-trivial deformations.

We are still missing tori with $f$- and $R$-flux, as they should arise from the T-duality chain
\begin{equation}
  H_{ijk} \stackrel{\mathcal{T}_i}{\longrightarrow} f^{i}_{jk}
  \stackrel{\mathcal{T}_j}{\longrightarrow} Q_{k}^{ij}
  \stackrel{\mathcal{T}_k}{\longrightarrow} R^{ijk}
\end{equation}
where $\mathcal{T}_i$ denotes a T-duality transformation along the coordinate $x^i$. In order to obtain the former, we have to remember that the deformations we performed above only contain SO($3$)$\times$SO($3$) group elements with determinate $1$. However, from DFT we know that T-duality transformations are actually covered by O($d$)$\times$O($d$) transformations. Thus, we have to add an additional element $\mathcal{T}_{123}$ which acts as
\begin{align}
  t_1 & \rightarrow t^1 & 
  t_2 & \rightarrow t^2 & 
  t_3 & \rightarrow t^3 & 
  t^1 & \rightarrow t_1 & 
  t^2 & \rightarrow t_2 & 
  t^3 & \rightarrow t_3
\end{align}
on the generators of the Lie algebra $\mathfrak{g}$ and has determinate $-1$. With this element we can build the additional deformations
\begin{equation}
  \mathcal{T}_1 = \mathcal{T}_{23} \mathcal{T}_{123}\,,\quad
  \mathcal{T}_2 = \mathcal{T}_{13} \mathcal{T}_{123} \quad \text{and} \quad
  \mathcal{T}_3 = \mathcal{T}_{12} \mathcal{T}_{123}\,. 
\end{equation}
All of them give rise to an abelian subalgebra $\mathfrak{h}$. For $\mathcal{T}_{123}$ alone the resulting $\mathfrak{h}$ is not a subalgebra anymore. Meaning, our method does not produce a SC solution for the torus with $R$-flux.

\subsubsection*{Torus with $f$-Flux}
After the deformation 
\begin{align}
  \mathcal{T}_3: &&
  t_1 & \rightarrow t_1 & 
  t_2 & \rightarrow t_2 & 
  t_3 & \rightarrow t^3 &
  t^1 & \rightarrow t^1 &
  t^2 & \rightarrow t^2 & 
  t^3 & \rightarrow t_3
\end{align}
of the Lie algebra \eqref{eqn:LiealgHtorus}, the subalgebra $\mathfrak{h}$ is still abelian and the corresponding connection $A$ vanishes without switching on a $B$-field. For the components of the background generalized vielbein, we now obtain
\begin{equation}\label{eqn:EAiFflux}
  E_{a i} = \mathbf{h} \begin{pmatrix}
    0 & \tilde x^3 & 0 \\ - \tilde x^3 & 0 & 0 \\ 0 & 0 & 0
  \end{pmatrix}\,, \quad
  E^a{}_i = \begin{pmatrix}
    1 & 0 & 0 \\ 0 & 1 & 0 \\ \mathbf{h} x^2 & 0 & 1
  \end{pmatrix} \quad \text{and} \quad
  E^a{}_{\tilde i} = \begin{pmatrix}
    1 & 0 & 0 \\ 0 & 1 & 0 \\ 0 & 0 & 1
  \end{pmatrix}\,.
\end{equation}
Since the vielbein $E^a{}_i$ is not trivial, there exists a non-vanishing geometric flux
\begin{equation}
  f_{ab}{}^c = 2 E_{[a}{}^i \partial_i E_{b]}{}^j E^c{}_j\,.
\end{equation}
We rewrite it in terms of three different two-forms $f^a = \frac{1}{2} f_{bc}{}^a E^b{}_i E^c{}_j d x^i \wedge d x^j$ and notice that the only non-vanishing contribution is given by
\begin{equation}
  f^3 = \mathbf{h} \, d x^1 \wedge d x^2 = c_1
\end{equation}
which represents the first Chern class of a circle bundle over the base $T^2$ in the directions $x^1$ and $x^2$. The twist of the generalized Lie derivative on the generalized tangent bundle vanishes and we have no $H$-flux contribution from the frame field $\hat E_A$ either (as expected for the twisted torus). Hence, T-duality has exchanged the role of 
\begin{equation}
  \int_{T^2} c_1 \leftrightarrow \int_{T^3} H
\end{equation}
where $H$ denotes the non-trivial $H$-flux \eqref{eqn:HtorusH} in the T-dual frame. This result is in perfect agreement with the literature about T-duality for non-trivially fibered circle bundles \cite{Bouwknegt:2003vb,Hull:2006qs}. Our initial conjecture that different solutions of the SC lead to T-dual GGs is confirmed. It is very important to keep in mind that we do not have to choose any generalized metric, in order to perform these calculations. Because we are still on the same group manifold the two solutions~\eqref{eqn:EAiHflux} and \eqref{eqn:EAiFflux} are related to each other by the change of coordinates
\begin{align}\label{eqn:trafoH->f}
  x^1 &\rightarrow x^1 &
  \tilde x^1 &\rightarrow -\tilde x^1 - \frac{\mathbf{h}}2 x^2 \tilde x^3  \nonumber \\
  x^2 &\rightarrow x^2 &
  \tilde x^2 &\rightarrow -\tilde x^2 + \frac{\mathbf{h}}2 x^1 \tilde x^3  \nonumber \\
  x^3 &\rightarrow \tilde x^3 & 
  \tilde x^3 &\rightarrow - x^3 + \frac{\mathbf{h}}2 x^1 x^2 \,.
\end{align}
All fields on the doubled space may only depend on the variables $x^1,\,x^2,\,x^3$. Otherwise they would violate the SC. If there exist fields which only depend on $x^1,\,x^2$, they solve it before and after the coordinate transformations \eqref{eqn:trafoH->f}. For these fields we can freely choose a duality frame. This observation is in perfect agreement with the common fact that T-duality is only allowed along isometries of the background. After identifying T-duality transformations with coordinate changes (diffeomorphisms), it becomes clear why the covariance of \DFTwzw{} under standard diffeomorphisms is so important: It make T-duality manifest.

\subsubsection*{Torus with $Q$-Flux}
Finally, we perform the deformation
\begin{align}
  \mathcal{T}_{23}: &&
  t_1 & \rightarrow t_1 & 
  t_2 & \rightarrow t^2 & 
  t_3 & \rightarrow t^3 &
  t^1 & \rightarrow t^1 &
  t^2 & \rightarrow t_2 & 
  t^3 & \rightarrow t_3
\end{align}
in order to obtain a torus with $Q$-flux. As for the twisted torus, the connection $A$ is flat for $B=0$. We repeat all the steps outlined above and obtain a trivial generalized frame $\hat E_A$. From \eqref{eqn:scFABC} it follows directly that the twist $\mathcal{F}_{A}{}^C = F_{AB}{}^C$ is equivalent to the structure coefficient of the Lie algebra $\mathfrak{g}$. Its non-vanishing contribution is
\begin{equation}
  Q = \mathbf{h} \, \partial_2 \wedge \partial_3 \wedge d x^1
\end{equation}
As expected, we obtain a background with $Q$-flux which can be translated into the Poisson algebra
\begin{equation}
  \{ x^i , x^j \} = Q^{ij}_k x^k \quad \text{resulting in} \quad
  \{ x^2 , x^3 \} = \mathbf{h} x^1
\end{equation}
governing the coordinates of the physical subspace $M$ \cite{Lust:2010iy,Andriot:2012an,Blumenhagen:2012pc,Condeescu:2012sp,Bakas:2015gia}. Taking this identification into account, we conclude that we obtain a non-commutative two-torus ($x^2$, $x^3$) which is fibred over a circle with the coordinate $x^1$. The same conclusion was drawn in \cite{Mathai:2004qq}. Again, this frame can be related to the torus with $H$-flux by the coordinate transformation
\begin{align}
  x^1 &\rightarrow x^1 &
  \tilde x^1 &\rightarrow -\tilde x^1 - \frac{\mathbf{h}}2 \tilde x^2 \tilde x^3  \nonumber \\
  x^2 &\rightarrow \tilde x^2 &
  \tilde x^2 &\rightarrow -x^2 + \frac{\mathbf{h}}2 x^1 \tilde x^3  \nonumber \\
  x^3 &\rightarrow \tilde x^3 &
  \tilde x^3 &\rightarrow - x^3 - \frac{\mathbf{h}}2 x^1 \tilde x^2 \,.
\end{align}

Because the generalized frame field \eqref{eqn:genframe&dual} lacks a contribution from an anti-symmetric bivector $\beta^{ij}$, it cannot implement $Q$-flux. However, it would be interesting to extend the presented formalism to include it. In this paper, we start from a Lie Group with the corresponding Lie algebra and obtain a Courant algebroid. It can be rewritten in terms of two Lie algebroids fulfilling a compatibility condition. One for the tangent and one for the cotangent bundle. They give rise to a Lie bi-algebroid with the structure of a Courant algebroid (see \cite{Deser:2013pra} for a review). For the torus with  $H$- and $f$-flux, the Lie algebroid of the tangent bundle is non-trivial while the one for the cotangent bundle is trivial. For the $Q$- and $R$-flux frames, this situation is flipped. Thus, our present formalism seems to be unable to produce non-trivial Lie algebroids for the cotangent bundle. This is not surprising because we implemented the Lie algebra $\mathfrak{g}$ on the tangent bundle of the group manifold only. A generalization would consider the dual Lie algebra $\mathfrak{g}^*$, too. Combining it with $\mathfrak{g}$ one could obtain a Lie bi-algebra corresponding to a Poisson-Lie group. Starting from this generalized structure instead of a Lie group only, it could be possible to incorporate $Q$- and $R$-fluxes on the same footing as $H$- and $f$-flux. However, this modification is beyond the scope of this paper.

\subsection{Three-Sphere with \texorpdfstring{$H$}{H}-Flux}\label{sec:S3H}
To describe a $S^3$ with $\mathbf{h}$ units of $H$-flux, we consider the Lie algebra $\mathfrak{g}=\mathfrak{so}(4)$ defined by the non-vanishing commutators
\begin{equation}\label{eqn:so4}
  [ t_a, t_b ] = \sqrt{\frac2{\mathbf{h}}} \epsilon_{abc} t^c\,, \quad
  [ t_a, t^b ] = \sqrt{\frac2{\mathbf{h}}} \epsilon_a{}^{bc} t_c \quad \text{and} \quad
  [ t^a, t^b ] = \sqrt{\frac2{\mathbf{h}}} \epsilon^{ab}{}_c t^c\,.
\end{equation}
We identify the maximal isotropic subalgebra $\mathfrak{h}=\mathfrak{so}(3)$ and assign the remaining generators $t_a$ to $\mathfrak{m}$. These two subsets form a symmetric pair. Hence, the technique to find a flat connection from section~\ref{sec:linkspinoromega} applies to this example.

For the following discussion, it is convenient to remember the isomorphism
\begin{equation}\label{eqn:so4su2su2}
  \text{SO($4$)} \cong \text{SU($2$)} \times \text{SU($2$)}/ \mathbb{Z}_2\,.
\end{equation}
In order to implement it, take a point in $\mathbb{R}^4$ with coordinates $y^0$, $\dots$, $y^3$ and write it  in terms of the 2$\times$2 matrix
\begin{equation}
  Y = \mathbf{1} y^0 - i \sigma_j y^j
\end{equation}
with $\sigma_j$, $j$=1,2,3\,, denoting the Pauli-matrices. An arbitrary SO($4$) action on this point can be rewritten as
\begin{equation}
  Y \rightarrow g_L Y g_R^{-1} \quad \text{with} \quad g_L,\,g_R\in\text{SU($2$)}\,.
\end{equation}
Because the SU($2$)$\times$SU($2$) elements $(g_L,g_R)$ and $(-g_L,-g_R)$ mediate the same transformation, one has to mod out $\mathbb{Z}_2$ in \eqref{eqn:so4su2su2}. Furthermore, assume that $g_R$ and $g_L$ are generated by
\begin{equation}
  t_{L\,a} = - \frac{i}{\sqrt{2 \mathbf{h}}} \sigma_a \quad \text{and} \quad t_{R\,a} = \frac{i}{\sqrt{2 \mathbf{h}}} \sigma_a\,.
\end{equation}
These generators are governed by the non-vanishing commutator relations
\begin{equation}
  [t_{L\,a}, t_{L\,b}] = \sqrt{\frac2{\mathbf{h}}} \epsilon_{ab}{}^c t_{L\,c} \quad \text{and} \quad
  [t_{R\,a}, t_{R\,b}] = - \sqrt{\frac2{\mathbf{h}}} \epsilon_{ab}{}^c t_{R\,c}\,.
\end{equation}
They reproduce the algebra used in appendix A of \cite{Blumenhagen:2014gva} to discuss \DFTwzw{} for the SU($2$) WZW-model. This representation is well suited make contact with the underlying CFT description of the target space. By combining
\begin{equation}
  t_a = t_{L\,a} + t_{R\,a} \quad \text{and} \quad t^a = t_{L\,a} - t_{R\,a}\,,
\end{equation}
we recover the non-vanishing commutators in \eqref{eqn:so4}.

We choose a coset representative $m$ of the form \eqref{eqn:formsol}. As shown in section~\ref{sec:linkspinoromega}, it automatically solves of the SC. Applying the isomorphism \eqref{eqn:so4su2su2}, we write it as
\begin{equation}\label{eqn:mS3}
	m = (g_m, g_m^{-1}) \quad \text{with} \quad g_m \in \text{SU($2$)}\,.
\end{equation}
How is $m$ related to a three-sphere? From an abstract point of view, the isomorphism $g_m\ni$SU($2$) $\cong S^3$ is well known. Here, we want to make this connection manifest and derive a map from points on the three-sphere to coset representatives $m$. To this end, we first expand 
\begin{equation}
  g_m = \mathbf{1} x^0 - i \sigma_j x^j 
\end{equation}
in the same way as $Y$. Note that the additional constraint $(x^0)^2 + (x^1)^2 + (x^2)^2 + (x^3)^2 = 1$ holds, because $g_m \in$SU($2$). Applying $m$ to the unit vector represented by $Y=\mathbf{1}$ gives rise to
\begin{equation}\label{eqn:yi(xi)}
  Y = g_m g_m \quad \text{or equivalently} \quad
    y^0 = 2 (x^0)^2 - 1 \quad \text{and} \quad
    y^i = 2 x^0 x^i \,.
\end{equation}
If we choose hyperspherical coordinates
\begin{align}\label{eqn:hypershere}
  y^0 & = \cos \phi^1                &  y^2 & = \sin \phi^1 \sin \phi^2 \cos \phi^3 \nonumber \\
  y^1 & = \sin \phi^1 \cos \phi^2    &  y^3 & = \sin \phi^1 \sin \phi^2 \sin \phi^3
\end{align}
with
\begin{equation}
  0 \le \phi^1 \le \pi \,, \quad
	0 \le \phi^2 \le \pi  \quad \text{and} \quad
	0 \le \phi^3 \le 2\pi
\end{equation}
for a unit $S^3$ embedded in $\mathbb{R}^4$, we find the corresponding coset representative \eqref{eqn:mS3} with
\begin{equation}\label{eqn:hyperspheregm}
  g_m = \mathbf{1} \cos \frac{\phi^1}2 - i \sigma_1 \sin \frac{\phi^1}2 \cos \phi^2 -
    i \sigma_2 \sin \frac{\phi^1}2 \sin \phi^2 \cos \phi^3 - 
    i \sigma_3 \sin \frac{\phi^1}2 \sin \phi^2 \sin \phi^3 
\end{equation}
by inverting the relations \eqref{eqn:yi(xi)}.

Starting from the explicit parameterization \eqref{eqn:hyperspheregm} of the coset representative \eqref{eqn:mS3}, we calculate the components
\begin{equation}\label{eqn:EAiS3Hflux}
  E^a{}_i = \sqrt{\frac{\mathbf{h}}2} \begin{pmatrix}
    c_2  & - s_1 s_2  & 0 \\ 
    s_2 c_3 & s_1 c_2 c_3  & - s_1 s_2 s_3 \\
    s_2 s_3  & s_1 c_2 s_3  & \phantom{-} s_1 s_2 c_3
	\end{pmatrix} \quad \text{with} \quad
  c_i = \cos \phi^i \quad s_i = \sin \phi^i 
\end{equation}
and
\begin{equation}
  B = - \mathbf{h} \sin^2 \frac{\phi_1}2 \sin \phi^1 \sin \phi^2 \, d \phi^2 \wedge d \phi^3
\end{equation}
of the frame field \eqref{eqn:genframe&dual}. Its $B$-field gives rise to a non-vanishing $H$-flux
\begin{equation}
  H_{\hat E} = d B = - \mathbf{h} (1 + 2 \cos \phi^1) \sin^2 \frac{\phi^1}2 \sin \phi_2 \, d \phi^1 \wedge d \phi^2 \wedge d \phi^3 \,.
\end{equation}
However, $H_{\hat E}$ is trivial in cohomology because the integral
\begin{equation}
  [ H_{\hat E} ] = \frac1{4 \pi^2} \int_M H_{\hat E} = 0 
\end{equation}
over $M$=$S^3$ vanishes. Still, the total $H$-flux of the background is not trivial because there is another contribution for the structure coefficients of the Lie algebra $\mathfrak{g}$. It reads
\begin{equation}
  H = 2 \mathbf{h} \sin^2 \frac{\phi^1}2 \sin \phi^2 \, d x^1 \wedge d x^2 \wedge d x^3
\end{equation}
and integrates, after an appropriate normalization, to the cohomology class
\begin{equation}
  [ H ] = \frac1{4 \pi^2} \int_M H = \mathbf{h}\,.
\end{equation}
Indeed, we have a $S^3$ carrying $\mathbf{h}$ units of $H$-flux. The topology of the SC solution presented here is characterized by the splitting of this flux 
\begin{equation}
  [ H_{\hat E} ] = 0 \quad \text{and} \quad [ H_{\mathcal{F}} ] = \mathbf{h}
\end{equation}
between the frame field and the twist of the generalized Lie derivative. Note that this result is different from the one for the torus with $H$-flux in \eqref{eqn:classesTwithH}. We can explain it by taking a closer look at $H$=SO($3$)-principal bundles over three-spheres. First, split the $S^3$ into two overlapping patches $U_{\mathrm{N}}$/$U_{\mathrm{S}}$ covering its north/south pole. On each one there exists a local section $\sigma_{\mathrm{N}}$/$\sigma_{\mathrm{S}}$. Both are patched together along a two-sphere $S^2 = U_{\mathrm{N}} \cap U_{\mathrm{S}}$ on the equator. Topologically distinct transition functions
\begin{equation}
  t_{\mathrm{NS}} : S^2 \rightarrow \text{SO($3$)}
\end{equation}
are classified by the second homotopy group $\pi_2($SO($3$)$)$ which is trivial. Thus, the bundle always admits a global section and is trivial. At the same time, the resulting generalized tangent bundle $T^* M \oplus T M$ is not twisted and all non-trivial $H$-flux has to be implemented as a twist of the generalized Lie derivative.

Let us finally discuss an alternative SC solution on $G$. The subalgebra $\mathfrak{h}=\mathfrak{so}(3)$ is simple and does not admit any deformations. So we cannot apply the technique outlined in section~\ref{sec:deformations}. Still, we can modify the physical subspace $M$ which arises as a solution of the SC to reproduce the results about T-duality on the $S^3$ with $H$-flux \cite{Bouwknegt:2003vb}. To this end, consider the double coset
\begin{equation}
  M = \Gamma \setminus \text{SO($4$)} / \text{SO($3$)}
\end{equation}
where $\Gamma$ is a freely acting, discrete subgroup of SO($4$). Only discrete subgroups are allowed because they do not change the dimension of $M$. Here, we choose $\Gamma = \mathbb{Z}_n$. Its elements $k_l$, $l=0,\dots,n-1$, are embedded by
\begin{equation}
  k_l = \Big( e,\, \exp \frac{2\pi i l \sigma_3}n \Big)
\end{equation}
into SO($4$) and have the action
\begin{equation}
  z^1 \rightarrow e^{- 2\pi i l / n} z^1
    \quad \text{and} \quad
  z^2 \rightarrow e^{- 2\pi i l / n} z^2
    \quad \text{with} \quad
  z^1 = y^0 + i y^3 \quad z^2 = y^2 + i y ^1
\end{equation}
on the coordinates of $\mathbb{R}^4$. By identifying all points on the $S^3$ which are connected by the this action, a lens space $M$=L$(n, 1)$ arises as physical submanifold. Clearly this change of the physical manifold $M$ is not just a coordinate transformation as for the torus with $H$-flux in the last subsection. So what happens to an arbitrary function $f$ on the $S^3$ after modding out the discrete subgroup $\mathbb{Z}_n$? To answer this question, the hyperspherical coordinates in \eqref{eqn:hypershere} are not the best choice. Instead we switch to Euler angles
\begin{equation}
  z^1 = e^{ - i \xi / 2} \cos \frac\theta2 \quad \text{and} \quad
  z^2 = - e^{i (\phi - \xi/2 )} \sin \frac\theta2
\end{equation}
with
\begin{equation}
  0 \le \theta \le \pi\,, \quad
	0 \le \phi \le 2 \pi \quad \text{and} \quad
  0 \le \xi \le 4 \pi\,.
\end{equation}
In these coordinates the identification above only affects $\xi$ and we have to simply restrict its range \begin{equation}
  0 \le \xi \le \frac{4 \pi}n
\end{equation}
to obtain the lens space L$(n, 1)$. Single-valued functions $f(\theta, \phi, \xi)$ on the $S^3$ are in general multi-valued on L$(n, 1)$. However, we can expand them in a set of functions
\begin{equation}
  f(\theta, \phi, \xi) = \sum\limits_{w=-n+1}^{n-1} f^w(\theta, \phi, \xi) e^{2\pi i \xi w/n}
\end{equation}
which are single-valued on the lens space. These modes are very similar to winding modes on a circle. But there is only a finite number of them. That is why they cannot be associated to an additional direction with a winding coordinate. From a CFT point of view, the modes $f^w$ correspond to the twisted sector of the level $\mathbf{h}$ WZW-model orbifold \cite{Giddings:1993wn,Maldacena:2001ky}
\begin{equation}\label{eqn:orbifoldcft}
  \text{SU($2$)$_{\mathrm{L}}$} \times \frac{\text{SU($2$)$_{\mathrm{R}}$}}{%
    \mathbb{Z}_n}\,.
\end{equation}
A particularly important function is the spinor $|\Lambda\rangle$ which implements the SC solution on $G$. It is not single-valued on L$(n,1)$ and, similar to a scalar function, we can expand it into different winding contributions. The linear constraint \eqref{eqn:sclinear} is a spinor constraint. Thus, it decompose into $n$ different constraints, one for each sector with a specific winding number. In the orbifold CFT \eqref{eqn:orbifoldcft} there is a similar effect: Level matching depends on the winding number of the twisted sector \cite{Giddings:1993wn}. This two observations are closely related because in \DFTwzw{} the SC is a direct consequence of level matching.

Here, we are only interested in the $H$-flux of the untwisted sector $w=0$. Taking into account the structure coefficients \eqref{eqn:so4}, it is given by the equation
\begin{equation}
  H = \sqrt{\frac{2}{\mathbf{h}}} \langle \Lambda | \Lambda \rangle d V
\end{equation}
with
\begin{equation}
  dV = \det (E^a{}_i) \, d \theta \wedge d \phi \wedge d \xi
    \quad \text{and} \quad
  \langle \Lambda | \Lambda \rangle = 1 - \frac12 \Tr (B_{ab})^2\,.
\end{equation}
After calculating the $B$-field and the vielbein $E^a{}_i$ in Euler angles, we obtain
\begin{equation}
  dV = \left( \frac{\mathbf{h}}{2} \right)^{3/2} \frac18 \sin \theta \, d \theta \wedge d \phi \wedge d \xi
    \quad \text{and} \quad
  \langle \Lambda | \Lambda \rangle = \frac2{1 + \cos \frac\theta2 \cos \frac\xi2}\,.
\end{equation}
As expected, $\Lambda$ and therewith its absolute value squared are not single valued on L$(n, 1)$. In order to get a simple expression for it in the $w=0$ sector, we further assume that $n = \mathbf{h}$. In this case, we only have to evaluate the zero mode contribution
\begin{equation}
  \langle \Lambda | \Lambda \rangle^0 = \frac1{4\pi} \int_0^{4\pi} \langle \Lambda | \Lambda \rangle = \frac{2}{\sin \frac\theta2}\,.
\end{equation}
It gives rise to
\begin{equation}
  H^0 = \frac{\mathbf{h} \sin\theta}{8 \sin\frac\theta2} \, d \theta \wedge d \phi \wedge d \xi
\end{equation}
with the cohomology class
\begin{equation}
  [ H^0 ] = \frac1{4 \pi^2} \int_M H^0 = 1\,.
\end{equation}
Hence, we reproduce the topology change induces by T-duality along the Hopf-fiber of a $S^3$ with $\mathbf{h}$ units of $H$-flux. In agreement with \cite{Bouwknegt:2003vb}, the dual space is a lens space L$(\mathbf{h}, 1)$ with one unit of $H$-flux.

An extensive discussion of the doubled geometry of WZW-models is given in \cite{Schulz:2011ye}. As here, the doubled space is treated as a group manifold and a projection $\pi: G \rightarrow G/H$ is used to obtain the physical subspace. Still, we cannot reproduce the mechanism for abelian T-duality outlined there. Our results suggest that T-duality does not simply exchange a winding and a momentum circle over a two-sphere base. Its action is more subtle due to the finite number of winding modes.

\section{Outlook}\label{sec:outlook}
In this paper we show how the SC of \DFTwzw{} is solved and how its solutions are related to T-duality. Doing so, only the topology of the group manifold $G$ enters the discussion. We do not need to fix a generalized metric and still reproduce topology changes induces by T-duality \cite{Bouwknegt:2003vb,Hull:2006qs}. There are still various open questions:

\begin{itemize}
  \item The examples we present do not include backgrounds which are not T-dual to geometric one. These have geometric and non-geometric fluxes with common directions. $R$-flux presents an obstruction to find a maximal isotropic subgroup \cite{Grana:2008yw,Hull:2009sg}, but there are also examples with $Q$-flux like the double elliptic setup in \cite{Hassler:2014sba} which can be treated with the techniques introduces here. It would be very interesting to study whether such backgrounds admit SC solutions or not.
  \item Solitonic objects like branes and monopoles are essential in studying string theory. They are characterized by topological charges which change under T-duality. An illustrative example is the NS5-brane. Take $\mathbf{h}$ of them sitting at a point in the transverse space $\mathbb{R}^3\times S^1$ and a two-sphere $S^2$ which encompasses the stack in $\mathbb{R}^3$. The number of branes follows directly from the $H$-flux
    \begin{equation}
      \frac{1}{4\pi^2} \int_{S^2\times S^1} H = \mathbf{h}
    \end{equation}
    because each NS5-brane contributes one unit of $H$-flux. After applying T-duality along the $S^1$, it becomes non-trivial fibered over the $S^2$ and obtains a non-vanishing first Chern class $[c_1]=\mathbf{h}$. The resulting space is formed by $\mathbf{h}$ Kaluza-Klein monopoles \cite{Bouwknegt:2003vb}. In the same vein, the example $S^3$ with $H$-flux in section~\ref{sec:S3H} is equivalent to a configuration with one KK-monopole and $\mathbf{h}$ NS5-branes. Studying these objects and also their non-geometric counter parts \cite{deBoer:2012ma,Hassler:2013wsa} is a prominent application of DFT \cite{Berkeley:2014nza,Berman:2014jsa,Bakhmatov:2016kfn}. Currently, only local solutions in terms of an explicit generalized metric are know \cite{Berman:2014jsa}. However, there is also a very interesting global perspective as the NS5-brane/KK-monopole example above shows. We hope that \DFTwzw{} combined with the techniques present in this paper facilitate further progress in this direction.
  \item By solving the SC, we obtain exact Courant algebroids. There is a lot to say about them and their connection to DFT \cite{Deser:2014mxa,Deser:2016qkw}. Here, we only want to highlight that each Courant algebroid is linked to a Courant $\sigma$-model which plays an important role in understanding the non-associative spacetime in $R$-flux backgrounds \cite{Lust:2010iy,Blumenhagen:2010hj,Blumenhagen:2011ph,Mylonas:2012pg,Bakas:2013jwa,Blumenhagen:2013zpa}. On the other hand a $S^3$ with $H$-flux is automatically non-associative for a finite among of flux. To see this, remember that the corresponding $\sigma$-model gives rise a SU($2$)$\times$SU($2$) Ka\v{c}-Moody algebra at level $\mathbf{h}$. Primary fields are translated to a set of orthogonal functions on the physical space $M=S^3$. Their angular momentum cannot exceed the level. After multiplying two of them, one can in general exceed the maximally allowed angular momentum. Therefore a projector is required. It renders the algebra of functions on $M$ non-associative \cite{Ramgoolam:2001zx}. For $\mathbf{h}\to \infty$ this effect vanishes. It is directly related to $\alpha'$ corrections because in \DFTwzw{} we have an expansion in $1/\mathbf{h}$ instead of $\alpha'$ \cite{Blumenhagen:2014gva}. Now the speculative part of the story: T-duality along all directions of the target space should inverts the worldsheet coupling $1/\mathbf{h}\rightarrow \mathbf{h}$. As a result there is a transition from a large, semiclassical $S^3$ to a very small, strongly non-associative background. We have to face two major challenges in trying to test this idea. First, $\alpha'$ corrections are not worked out in \DFTwzw{}. Furthermore, we are not able to find solutions to the SC with non-trivial $R$-flux. We hope that in the future enough progress is made to over come these challenges. Additional hints in the right direction are given by \cite{Blumenhagen:2011ph,Blumenhagen:2010hj,Blumenhagen:2013zpa}.
\end{itemize}

\acknowledgments
I would like to thank
David Berman,
Pascal de Bosque,
Martin Cederwall,
Jonathan Heckman,
Olaf Hohm,
Chris Hull,
Dieter L\"ust,
Emanuel Malek and
Daniel Waldram
for helpful discussions. Furthermore, I would like to thank the ITS at the City University of New York and the theory group at Columbia University for their hospitality during parts of this project. My work is supported by the NSF CAREER grant PHY-1452037. I also acknowledge support from the Bahnson Fund at UNC Chapel Hill as well as the R. J. Reynolds Industries, Inc. Junior Faculty Development Award from the Office of the Executive Vice Chancellor and Provost at UNC Chapel Hill and from the NSF grant PHY-1620311.

\appendix
\section{Lie Algebra Cohomology}\label{app:liealgebracohomology}
A powerful tool to classify deformations of Lie algebras is Lie algebra cohomology. Its main objects are $r$-cochains, which form the set $\Omega^r(\mathfrak{g}, V)$. They represent totally anti-symmetric, multi-linear maps
\begin{equation}
  \omega : \underbrace{\mathfrak{g} \times \dots \times \mathfrak{g}}_{\text{$r$ times}} \rightarrow V \quad \omega \in \Omega^r(\mathfrak{g}, V)
\end{equation}
from the tensor product of $r$ times the Lie algebra $\mathfrak{g}$ to a vector space $V$, which carries a representation
\begin{equation}
  \pi: \mathfrak{g} \rightarrow \mathfrak{gl}(V) \quad \text{with} \quad
  [ \pi(X), \pi(Y)] = \pi ([X,Y]) \quad \forall\, X,\,Y \in \mathfrak{g}
\end{equation}
of the Lie algebra. A $r$-cochain is mapped to a $r+1$-cochain by the coboundary operator
\begin{equation}
  d: \Omega^r(\mathfrak{g}, V) \rightarrow \Omega^{r+1}(\mathfrak{g}, V)
\end{equation}
which is defined as
\begin{align}\label{eqn:defcoboundary_d}
  d \omega( X_{i_1}, \dots, X_{i_{r+1}}) = &\sum\limits_{1\le k < l \le r+1}^{r+1} \omega([X_{i_k}, X_{i_l}], X_{i_1}, \dots, \hat{X}_{i_k}, \dots, \hat{X}_{i_l}, \dots, X_{i_{r+1}}) \nonumber \\
  &\,\,\, + \sum\limits_{k=1}^{r+1} (-1)^{k+1} \pi(X_{i_k}) \omega(X_{i_1}, \dots, \hat{X}_{i_k}, \dots , X_{i_{r+1}})\,.
\end{align}
Hatted $X_i$ are omitted from the arguments of the cochain. A convenient way to write $r$-cochains is in terms of basis one-forms $\theta^A \in \Omega^1(\mathfrak{g}, \mathds{R})$ with the defining property $\theta^A(t_B)=\delta^A_B$ as
\begin{equation}
  \omega = \frac{1}{r!} \omega_{{A_1} \dots A_r}{}^\alpha e_\alpha\, \theta^{A_1} \wedge \dots \wedge \theta^{A_r}\,.
\end{equation}
Here, the indices $\alpha$ run from one to the dimension of $V$ and $e_\alpha$ denotes a set of basis vectors spanning $V$. In this notation, \eqref{eqn:defcoboundary_d} translates into
\begin{equation}
  d \theta^A = -\frac{1}{2} f_{BC}{}^A \theta^B \wedge \theta^C
    \quad \text{and} \quad
  d e_\alpha = \pi(t_A) e_\alpha \theta^A\,,
\end{equation}
where $f_{AB}{}^C$ are the structure coefficients of $\mathfrak{g}$. In order to obtain the correct signs, one further has to remember the property of the exterior derivative
\begin{equation}
  d ( \alpha \wedge \beta ) = d \alpha \wedge \beta + (-1)^p \alpha \wedge d \beta 
   \quad \text{with} \quad
  \alpha \in \Omega^q(\mathfrak{g}, \mathds{R})\,, \quad 
  \beta  \in \Omega^p(\mathfrak{g}, \mathds{R})\,.
\end{equation}
Using the Jacobi identity of the Lie algebra, it is straightforward to show that
\begin{equation}
  d^2  = 0\,.
\end{equation}
Thus, $d$ gives rise a graded differential complex
\begin{equation}
  \cdots \xrightarrow{\rule{1em}{0pt}d\rule{1em}{0pt}} \Omega^{r-1}(\mathfrak{g}, V) \xrightarrow{\rule{1em}{0pt}d\rule{1em}{0pt}} \Omega^r(\mathfrak{g}, V) \xrightarrow{\rule{1em}{0pt}d\rule{1em}{0pt}} \Omega^{r+1}(\mathfrak{g}, V) \xrightarrow{\rule{1em}{0pt}d\rule{1em}{0pt}} \cdots
\end{equation}
which is called the Chevalley-Eilenberg complex of the algebra $\mathfrak{g}$ with coefficients in $V$. A $r$-cochain is called closed (or cocycle), if $d \omega = 0$ holds. A $r$-cocycle, which can be expressed in terms of a $(r-1)$-cochain as $\omega = d \alpha$, is called exact (or coboundary). Taking the quotient of $r$-cocycles and $r$-coboundaries, we obtain the Lie algebra cohomology
\begin{equation}
  H^r(\mathfrak{g}, V) = \frac{\text{ker} \, d: \Omega^r (\mathfrak{g}, V) \rightarrow \Omega^{r+1} (\mathfrak{g}, V)}{\text{im} \, d: \Omega^{r-1} (\mathfrak{g}, V) \rightarrow \Omega^r(\mathfrak{g},V)} \,.
\end{equation}

For infinitesimal deformations of Lie algebras, $H^2(\mathfrak{g}, \mathfrak{g})$ plays a central role. In order to see why, consider the deformed Lie bracket
\begin{equation}\label{eqn:deformedbracket}
  [ X, Y ]_\lambda = [X , Y] + \lambda c_1 (X, Y) + \lambda^2 c_2 (X, Y) + \lambda^3 c_3 (X, Y) + \cdots\,,
\end{equation}
where $c_1, c_2, c_3, \dots$ are elements of $\Omega^2(\mathfrak{g}, \mathfrak{g})$. Of course, this new bracket should not violate the Jacobi identity 
\begin{equation}\label{eqn:deformedjacobi}
  [ X, [ Y, Z ]_\lambda ]_\lambda + [ Z, [ X, Y ]_\lambda ]_\lambda + [ Y, [ Z, X ]_\lambda ]_\lambda = 0
    \quad\text{with}\quad
  X, Y, Z \in \mathfrak{g}\,.
\end{equation}
Thus, there arise several constraints on the two-cochains in the expansion \eqref{eqn:deformedbracket}. First, we only take into account terms up to linear order in $\lambda$, which yields that $c_1$ has to be a cocycle. If it is a coboundary, we can rewrite $c_1$ as $c_1 = d b_1$ and are able to describe deformed bracket
\begin{equation}
  [T(X), T(Y)]_\lambda = T([X, Y])
\end{equation}
in terms of the isomorphism
\begin{equation}
  T: \mathfrak{g} \rightarrow \mathfrak{g}
    \quad\text{with}\quad
  T(X) = X + \lambda b_1(X)
\end{equation}
on the Lie algebra. Such deformations are called trivial. All non-trivial, linear deformations are therefore elements of $H^2(\mathfrak{g}, \mathfrak{g})$.  All higher orders of the deformation are governed by the constraint
\begin{equation}\label{eqn:constrdeformation}
  d c_k = - \sum\limits_{i=1}^{k-1} [ c_i, c_k ]
\end{equation}
where the bracket denotes the Massey product
\begin{equation}
  [\alpha, \beta](X_1, \dots, X_{p+q-1}) = \alpha(\beta(X_1, \dots X_q), X_{q+1}, \dots, X_{q+p-1}) + \text{cyclic}
\end{equation}
for $\alpha \in \Omega^p(\mathfrak{g}, \mathfrak{g})$ and $\beta \in \Omega^q(\mathfrak{g}, \mathfrak{g})$. Keeping in mind that the Massey product of two cocycles is again a cocycle, the obstructions of solving the constraints \eqref{eqn:constrdeformation} are given by $H^3(\mathfrak{g}, \mathfrak{g})$.

\section{Identifications for the coset CSO($1,0,3,\mathbb{Z}$)\textbackslash CSO($1,0,3$)}%
\label{sec:modoutcso(103z)}
In this appendix we derive the identifications \eqref{eqn:identtwithH} which arise after modding out the discrete subgroup CSO($1,0,3,\mathbb{Z}$) from the Lie group CSO($1,0,3$). Let us start with a group element $g=m h$ which can be parameterized by the coordinates in \eqref{eqn:paramCSO(1,0,3)}. Following the appendix in \cite{Bosque:2015jda}, we write it in terms of the matrix
\begin{equation}
g = \left(
\begin{array}{ccccccccccccc}
 1 & 0 & 0 & 0 & 0 & 0 & 0 & 0 & 0 & 0 & 0 & 0 & x^1 \\
 0 & 1 & 0 & 0 & 0 & 0 & 0 & 0 & 0 & 0 & 0 & 0 & x^2 \\
 0 & 0 & 1 & 0 & 0 & 0 & 0 & 0 & 0 & 0 & 0 & 0 & x^3 \\
 x^1 & 0 & 0 & 1 & 0 & 0 & 0 & 0 & 0 & 0 & 0 & 0 & \frac{(x^1)^2}{2} \\
 x^2 & x^1 & 0 & 0 & 1 & 0 & 0 & 0 & 0 & 0 & 0 & 0 & x^1 x^2 \\
 x^3 & 0 & x^1 & 0 & 0 & 1 & 0 & 0 & 0 & 0 & 0 & 0 & x^1 x^3 \\
 0 & x^2 & 0 & 0 & 0 & 0 & 1 & 0 & 0 & 0 & 0 & 0 & \frac{(x^2)^2}{2} \\
 0 & x^3 & x^2 & 0 & 0 & 0 & 0 & 1 & 0 & 0 & 0 & 0 & x^2 x^3 \\
 0 & 0 & x^3 & 0 & 0 & 0 & 0 & 0 & 1 & 0 & 0 & 0 & \frac{(x^3)^2}{2} \\
 0 & -x^3 \mathbf{h} & 0 & 0 & 0 & 0 & 0 & 0 & 0 & 1 & 0 & 0 & \tilde x^1 \\
 x^3 \mathbf{h} & 0 & 0 & 0 & 0 & 0 & 0 & 0 & 0 & 0 & 1 & 0 &  \tilde x^2 \\
 -x^2 \mathbf{h} & 0 & 0 & 0 & 0 & 0 & 0 & 0 & 0 & 0 & 0 & 1 & \tilde x^3 \\
 0 & 0 & 0 & 0 & 0 & 0 & 0 & 0 & 0 & 0 & 0 & 0 & 1 \\
\end{array}
\right)
\end{equation}
where $\mathbf{h}$ is an integer denoting the number of $H$-flux units in the background. Working with such large matrices is cumbersome. So we represent $g$ instead by the six tuple $(x^1\,,x^2\,,x^3\,,\tilde x^1\,,\tilde x^2\,, \tilde x^3)$. In this case, the group multiplication is given by
\begin{gather}
  (x^1\,,x^2\,,x^3\,,\tilde x^1\,,\tilde x^2\,, \tilde x^3) (y^1\,,y^2\,,y^3\,,\tilde y^1\,,\tilde y^2\,, \tilde y^3) = \nonumber\\ 
  (x^1+y^1\,,x^2+y^2\,,x^3+y^3\,,- x^3 y^2 \mathbf{h} + \tilde x^1 + \tilde y^1\,, x^3 y^1 \mathbf{h} + \tilde x^2 + \tilde y^2\,,- x^2 y^1 \mathbf{h} + \tilde x^3 + \tilde y^3)\,. \label{eqn:groupmult}
\end{gather}
Let us check that this indeed gives rise to a group. The identity elements is $e=(0\,,0\,,0\,,0\,,0\,,0)$ and fulfills
\begin{equation}
  g e = e g = g\,.
\end{equation}
Furthermore, there is the inverse element
\begin{equation}
  g^{-1}=(-x^1\,,-x^2\,,-x^3\,, -x^3 x^2 \mathbf{h} - \tilde x^1\,, x^3 x^1 \mathbf{h} - \tilde x^2\,, -x^2 x^1 \mathbf{h} - \tilde x^3)
\end{equation}
fulfilling
\begin{equation}
  g^{-1} g = g g^{-1} = e\,.
\end{equation}
Because $\mathbf{h}$ is an integer, the group multiplication \eqref{eqn:groupmult} does not only close over the real numbers, but also for $x^i$ and $\tilde x^{\tilde i}$ being integers. Thus CSO($1,0,3,\mathbb{Z}$) is a subgroup of CSO($1,0,3$) and we can mod it out by considering the right coset CSO($1,0,3,\mathbb{Z}$)\textbackslash CSO($1,0,3$). It gives rise to the equivalence relation
\begin{equation}
  g_1 \sim g_2 \quad \text{if and only if} \quad g_1 = k g_2 \quad \text{with} \quad
  g_1\,, g_2 \in \mathrm{CSO}(1,0,3) \quad \text{and} \quad
  k \in \mathrm{CSO}(1,0,3,\mathbb{Z})\,.
\end{equation}
After substituting $k=(n^1\,,n^2\,,n^3\,,\tilde n^1\,,\tilde n^2\,,\tilde n^3)$ with $n^i$, $\tilde n^{\tilde i} \in\mathbb{Z}$, we obtain the identifications
\begin{gather}
  (x^1\,,x^2\,,x^3\,,\tilde x^1\,,\tilde x^2\,, \tilde x^3) \sim \nonumber\\ 
  (x^1+n^1\,,x^2+n^2\,,x^3+n^3\,,-x^2 n^3 \mathbf{h} + \tilde x^1 + \tilde n^1\,,x^1 n^3 \mathbf{h} + \tilde x^2 + \tilde n^2\,,-x^1 n^2 \mathbf{h} + \tilde x^3 + \tilde n^3)
\end{gather}
from \eqref{eqn:groupmult}. Especially, we have
\begin{align}\label{eqn:identif1}
  (x^1\,,x^2\,,x^3\,,\tilde x^1\,,\tilde x^2\,, \tilde x^3) &\sim (x^1 + 1\,, x^2\,, x^3\,, \tilde x^1\,, \tilde x^2\,, \tilde x^3) \nonumber \\
  & \sim (x^1\,, x^2 + 1\,, x^3\,, \tilde x^1\,, \tilde x^2\,, \tilde x^3 - x^1 \mathbf{h}) \nonumber \\
  & \sim (x^1 \,, x^2\,, x^3 + 1\,, \tilde x^1 - x^2 \mathbf{h}\,, \tilde x^2 + x^1 \mathbf{h}\,, \tilde x^3)
\end{align}
for the physical coordinates and
\begin{align}
  (x^1\,,x^2\,,x^3\,,\tilde x^1\,,\tilde x^2\,, \tilde x^3) &\sim (x^1\,, x^2\,, x^3\,, \tilde x^1 + 1\,, \tilde x^2\,, \tilde x^3) \nonumber \\
  & \sim (x^1\,, x^2\,, x^3\,, \tilde x^1\,, \tilde x^2 + 1\,, \tilde x^3) \nonumber \\
  & \sim (x^1 \,, x^2\,, x^3\,, \tilde x^1\,,\tilde x^2\,, \tilde x^3 + 1) \label{eqn:identif6}
\end{align}
for the tilded ones. Taking into account this patching, the vielbein
\begin{equation}\label{eqn:genvielbeinT3H}
  E^A{}_I = \left(
\begin{array}{cccccc}
 1 & 0 & 0 & 0 & 0 & 0 \\
 0 & 1 & 0 & 0 & 0 & 0 \\
 0 & 0 & 1 & 0 & 0 & 0 \\
 0 & x^3 \mathbf{h} & 0 & 1 & 0 & 0 \\
 -x^3 \mathbf{h} & 0 & 0 & 0 & 1 & 0 \\
 x^2 \mathbf{h} & 0 & 0 & 0 & 0 & 1 \\
\end{array}
\right) \quad \text{or} \quad
\begin{aligned}
	E_1 &= d x^1 \\
	E_2 &= d x^2 \\
  E_3 &= d x^3 \\
  E^1 &= d \tilde x_1 + x^3 \mathbf{h} d x^2 = d (\tilde x_1 - x^2 \mathbf{h}) + (x^3 + 1) \mathbf{h} d x^2\\
  E^2 &= d \tilde x_2 - x^3 \mathbf{h} d x^1 = d (\tilde x_2 + x^1 \mathbf{h}) - (x^3 + 1) \mathbf{h} d x^1\\
  E^3 &= d \tilde x_3 + x^2 \mathbf{h} d x^1 = d (\tilde x_3 - x^1 \mathbf{h}) + (x^2 + 1) \mathbf{h} d x^1
\end{aligned}
\end{equation}
is globally well-defined.

\bibliography{literatur}

\providecommand{\href}[2]{#2}\begingroup\raggedright\begin{thebibliography}{10}

\bibitem{Bouwknegt:2003vb}
P.~Bouwknegt, J.~Evslin, and V.~Mathai, {\it {T}-duality: {T}opology change
  from {H}-flux},  {\em Commun. Math. Phys.} {\bf 249} (2004) 383--415,
  [\href{http://arxiv.org/abs/hep-th/0306062}{{\tt hep-th/0306062}}].

\bibitem{Hull:1994ys}
C.~M. Hull and P.~K. Townsend, {\it {U}nity of {S}uperstring {D}ualities},
  {\em Nucl. Phys.} {\bf B438} (1995) 109--137,
  [\href{http://arxiv.org/abs/hep-th/9410167}{{\tt hep-th/9410167}}].

\bibitem{Witten:1995ex}
E.~Witten, {\it {S}tring theory dynamics in various dimensions},  {\em Nucl.
  Phys.} {\bf B443} (1995) 85--126,
  [\href{http://arxiv.org/abs/hep-th/9503124}{{\tt hep-th/9503124}}].

\bibitem{Buscher:1987sk}
T.~Buscher, {\it {A} {S}ymmetry of the {S}tring {B}ackground {F}ield
  {E}quations},  {\em Phys.Lett.} {\bf B194} (1987) 59.

\bibitem{Siegel:1993th}
W.~Siegel, {\it {S}uperspace duality in low-energy superstrings},  {\em
  Phys.Rev.} {\bf D48} (1993) 2826--2837,
  [\href{http://arxiv.org/abs/hep-th/9305073}{{\tt hep-th/9305073}}].

\bibitem{Hull:2009mi}
C.~Hull and B.~Zwiebach, {\it {D}ouble {F}ield {T}heory},  {\em JHEP} {\bf
  0909} (2009) 099, [\href{http://arxiv.org/abs/0904.4664}{{\tt
  arXiv:0904.4664}}].

\bibitem{Hull:2009zb}
C.~Hull and B.~Zwiebach, {\it {T}he {G}auge algebra of double field theory and
  {C}ourant brackets},  {\em JHEP} {\bf 0909} (2009) 090,
  [\href{http://arxiv.org/abs/0908.1792}{{\tt arXiv:0908.1792}}].

\bibitem{Hohm:2010pp}
O.~Hohm, C.~Hull, and B.~Zwiebach, {\it {G}eneralized metric formulation of
  double field theory},  {\em JHEP} {\bf 1008} (2010) 008,
  [\href{http://arxiv.org/abs/1006.4823}{{\tt arXiv:1006.4823}}].

\bibitem{Aldazabal:2013sca}
G.~Aldazabal, D.~Marques, and C.~Nunez, {\it {D}ouble {F}ield {T}heory: {A}
  {P}edagogical {R}eview},  {\em Class.Quant.Grav.} {\bf 30} (2013) 163001,
  [\href{http://arxiv.org/abs/1305.1907}{{\tt arXiv:1305.1907}}].

\bibitem{Hohm:2013bwa}
O.~Hohm, D.~Lüst, and B.~Zwiebach, {\it {T}he {S}pacetime of {D}ouble {F}ield
  {T}heory: {R}eview, {R}emarks, and {O}utlook},  {\em Fortsch.Phys.} {\bf 61}
  (2013) 926--966, [\href{http://arxiv.org/abs/1309.2977}{{\tt
  arXiv:1309.2977}}].

\bibitem{Vaisman:2012ke}
I.~Vaisman, {\it {O}n the geometry of double field theory},  {\em J. Math.
  Phys.} {\bf 53} (2012) 033509, [\href{http://arxiv.org/abs/1203.0836}{{\tt
  arXiv:1203.0836}}].

\bibitem{Vaisman:2012px}
I.~Vaisman, {\it {T}owards a double field theory on para-{H}ermitian
  manifolds},  {\em J. Math. Phys.} {\bf 54} (2013) 123507,
  [\href{http://arxiv.org/abs/1209.0152}{{\tt arXiv:1209.0152}}].

\bibitem{Cederwall:2014kxa}
M.~Cederwall, {\it {T}he geometry behind double geometry},  {\em JHEP} {\bf 09}
  (2014) 070, [\href{http://arxiv.org/abs/1402.2513}{{\tt arXiv:1402.2513}}].

\bibitem{Cederwall:2014opa}
M.~Cederwall, {\it {T}-duality and non-geometric solutions from double
  geometry},  {\em Fortsch.Phys.} {\bf 62} (2014) 942,
  [\href{http://arxiv.org/abs/1409.4463}{{\tt arXiv:1409.4463}}].

\bibitem{Hohm:2012gk}
O.~Hohm and B.~Zwiebach, {\it {L}arge {G}auge {T}ransformations in {D}ouble
  {F}ield {T}heory},  {\em JHEP} {\bf 1302} (2013) 075,
  [\href{http://arxiv.org/abs/1207.4198}{{\tt arXiv:1207.4198}}].

\bibitem{Park:2013mpa}
J.-H. Park, {\it {C}omments on double field theory and diffeomorphisms},  {\em
  JHEP} {\bf 06} (2013) 098, [\href{http://arxiv.org/abs/1304.5946}{{\tt
  arXiv:1304.5946}}].

\bibitem{Berman:2014jba}
D.~S. Berman, M.~Cederwall, and M.~J. Perry, {\it {G}lobal aspects of double
  geometry},  {\em JHEP} {\bf 1409} (2014) 066,
  [\href{http://arxiv.org/abs/1401.1311}{{\tt arXiv:1401.1311}}].

\bibitem{Hull:2014mxa}
C.~M. Hull, {\it {F}inite {G}auge {T}ransformations and {G}eometry in {D}ouble
  {F}ield {T}heory},  {\em JHEP} {\bf 04} (2015) 109,
  [\href{http://arxiv.org/abs/1406.7794}{{\tt arXiv:1406.7794}}].

\bibitem{Naseer:2015tia}
U.~Naseer, {\it A note on large gauge transformations in double field theory},
  {\em JHEP} {\bf 06} (2015) 002, [\href{http://arxiv.org/abs/1504.05913}{{\tt
  arXiv:1504.05913}}].

\bibitem{Papadopoulos:2014mxa}
G.~Papadopoulos, {\it Seeking the balance: Patching double and exceptional
  field theories},  {\em JHEP} {\bf 10} (2014) 089,
  [\href{http://arxiv.org/abs/1402.2586}{{\tt arXiv:1402.2586}}].

\bibitem{Howe:2016ggg}
P.~S. Howe and G.~Papadopoulos, {\it {P}atching {DFT}, {T}-duality and
  {G}erbes},  {\em JHEP} {\bf 04} (2017) 074,
  [\href{http://arxiv.org/abs/1612.07968}{{\tt arXiv:1612.07968}}].

\bibitem{Alvarez:1993qi}
E.~Alvarez, L.~Alvarez-Gaume, J.~L.~F. Barbon, and Y.~Lozano, {\it {S}ome
  global aspects of duality in string theory},  {\em Nucl. Phys.} {\bf B415}
  (1994) 71--100, [\href{http://arxiv.org/abs/hep-th/9309039}{{\tt
  hep-th/9309039}}].

\bibitem{Hull:2006qs}
C.~M. Hull, {\it {G}lobal aspects of {T}-duality, gauged sigma models and
  {T}-folds},  {\em JHEP} {\bf 10} (2007) 057,
  [\href{http://arxiv.org/abs/hep-th/0604178}{{\tt hep-th/0604178}}].

\bibitem{Hull:2005hk}
C.~Hull and R.~Reid-Edwards, {\it {F}lux compactifications of string theory on
  twisted tori},  {\em Fortsch.Phys.} {\bf 57} (2009) 862--894,
  [\href{http://arxiv.org/abs/hep-th/0503114}{{\tt hep-th/0503114}}].

\bibitem{Hull:2007jy}
C.~M. Hull and R.~A. Reid-Edwards, {\it Gauge symmetry, t-duality and doubled
  geometry},  {\em JHEP} {\bf 08} (2008) 043,
  [\href{http://arxiv.org/abs/0711.4818}{{\tt arXiv:0711.4818}}].

\bibitem{Dall'Agata:2008qz}
G.~Dall'Agata and N.~Prezas, {\it Worldsheet theories for non-geometric string
  backgrounds},  {\em JHEP} {\bf 08} (2008) 088,
  [\href{http://arxiv.org/abs/0806.2003}{{\tt arXiv:0806.2003}}].

\bibitem{Hull:2009sg}
C.~Hull and R.~Reid-Edwards, {\it {N}on-geometric backgrounds, doubled geometry
  and generalised {T}-duality},  {\em JHEP} {\bf 0909} (2009) 014,
  [\href{http://arxiv.org/abs/0902.4032}{{\tt arXiv:0902.4032}}].

\bibitem{Scherk:1979zr}
J.~Scherk and J.~H. Schwarz, {\it {H}ow to {G}et {M}asses from {E}xtra
  {D}imensions},  {\em Nucl.Phys.} {\bf B153} (1979) 61--88.

\bibitem{Scherk:1978ta}
J.~Scherk and J.~H. Schwarz, {\it {S}pontaneous {B}reaking of {S}upersymmetry
  {T}hrough {D}imensional {R}eduction},  {\em Phys.Lett.} {\bf B82} (1979) 60.

\bibitem{Aldazabal:2011nj}
G.~Aldazabal, W.~Baron, D.~Marques, and C.~Nunez, {\it {T}he effective action
  of {D}ouble {F}ield {T}heory},  {\em JHEP} {\bf 1111} (2011) 052,
  [\href{http://arxiv.org/abs/1109.0290}{{\tt arXiv:1109.0290}}].

\bibitem{Geissbuhler:2011mx}
D.~Geissbuhler, {\it {D}ouble {F}ield {T}heory and {N}=4 {G}auged
  {S}upergravity},  {\em JHEP} {\bf 1111} (2011) 116,
  [\href{http://arxiv.org/abs/1109.4280}{{\tt arXiv:1109.4280}}].

\bibitem{Geissbuhler:2013uka}
D.~Geissbuhler, D.~Marques, C.~Nunez, and V.~Penas, {\it {E}xploring {D}ouble
  {F}ield {T}heory},  {\em JHEP} {\bf 1306} (2013) 101,
  [\href{http://arxiv.org/abs/1304.1472}{{\tt arXiv:1304.1472}}].

\bibitem{Blumenhagen:2014gva}
R.~Blumenhagen, F.~Hassler, and D.~L\"ust, {\it {D}ouble {F}ield {T}heory on
  {G}roup {M}anifolds},  {\em JHEP} {\bf 02} (2015) 001,
  [\href{http://arxiv.org/abs/1410.6374}{{\tt arXiv:1410.6374}}].

\bibitem{Blumenhagen:2015zma}
R.~Blumenhagen, P.~du~Bosque, F.~Hassler, and D.~L\"ust, {\it {G}eneralized
  {M}etric {F}ormulation of {D}ouble {F}ield {T}heory on {G}roup {M}anifolds},
  {\em JHEP} {\bf 08} (2015) 056, [\href{http://arxiv.org/abs/1502.02428}{{\tt
  arXiv:1502.02428}}].

\bibitem{Bosque:2015jda}
P.~du~Bosque, F.~Hassler, and D.~L\"ust, {\it {F}lux {F}ormulation of {DFT} on
  {G}roup {M}anifolds and {G}eneralized {S}cherk-{S}chwarz
  {C}ompactifications},  {\em JHEP} {\bf 02} (2016) 039,
  [\href{http://arxiv.org/abs/1509.04176}{{\tt arXiv:1509.04176}}].

\bibitem{Klimcik:1995ux}
C.~Klimcik and P.~Severa, {\it {D}ual {N}on-{A}belian {D}uality and the
  {D}rinfeld double},  {\em Phys. Lett.} {\bf B351} (1995) 455--462,
  [\href{http://arxiv.org/abs/hep-th/9502122}{{\tt hep-th/9502122}}].

\bibitem{Klimcik:1996nq}
C.~Klimcik and P.~Severa, {\it {N}on-{A}belian {M}omentum-{W}inding
  {E}xchange},  {\em Phys. Lett.} {\bf B383} (1996) 281--286,
  [\href{http://arxiv.org/abs/hep-th/9605212}{{\tt hep-th/9605212}}].

\bibitem{Severa:2016prq}
P.~\v{S}evera, {\it {P}oisson-{L}ie {T}-duality as a boundary phenomenon of
  {C}hern-{S}imons theory},  {\em JHEP} {\bf 05} (2016) 044,
  [\href{http://arxiv.org/abs/1602.05126}{{\tt arXiv:1602.05126}}].

\bibitem{Hitchin:2004ut}
N.~Hitchin, {\it {G}eneralized {C}alabi-{Y}au manifolds},  {\em
  Quart.J.Math.Oxford Ser.} {\bf 54} (2003) 281--308,
  [\href{http://arxiv.org/abs/math/0209099}{{\tt math/0209099}}].

\bibitem{Gualtieri:2003dx}
M.~Gualtieri, {\it {G}eneralized complex geometry},
  \href{http://arxiv.org/abs/math/0401221}{{\tt math/0401221}}.

\bibitem{Koerber:2010bx}
P.~Koerber, {\it {L}ectures on {G}eneralized {C}omplex {G}eometry for
  {P}hysicists},  {\em Fortsch.Phys.} {\bf 59} (2011) 169--242,
  [\href{http://arxiv.org/abs/1006.1536}{{\tt arXiv:1006.1536}}].

\bibitem{Shelton:2005cf}
J.~Shelton, W.~Taylor, and B.~Wecht, {\it {N}ongeometric flux
  compactifications},  {\em JHEP} {\bf 0510} (2005) 085,
  [\href{http://arxiv.org/abs/hep-th/0508133}{{\tt hep-th/0508133}}].

\bibitem{Dabholkar:2005ve}
A.~Dabholkar and C.~Hull, {\it {G}eneralised {T}-duality and non-geometric
  backgrounds},  {\em JHEP} {\bf 0605} (2006) 009,
  [\href{http://arxiv.org/abs/hep-th/0512005}{{\tt hep-th/0512005}}].

\bibitem{Schulz:2011ye}
M.~B. Schulz, {\it {T}-folds, doubled geometry, and the {SU}(2) {WZW} model},
  {\em JHEP} {\bf 1206} (2012) 158, [\href{http://arxiv.org/abs/1106.6291}{{\tt
  arXiv:1106.6291}}].

\bibitem{Giddings:1993wn}
S.~B. Giddings, J.~Polchinski, and A.~Strominger, {\it Four-dimensional black
  holes in string theory},  {\em Phys. Rev.} {\bf D48} (1993) 5784--5797,
  [\href{http://arxiv.org/abs/hep-th/9305083}{{\tt hep-th/9305083}}].

\bibitem{Maldacena:2001ky}
J.~M. Maldacena, G.~W. Moore, and N.~Seiberg, {\it {G}eometrical interpretation
  of {D}-branes in gauged {WZW} models},  {\em JHEP} {\bf 0107} (2001) 046,
  [\href{http://arxiv.org/abs/hep-th/0105038}{{\tt hep-th/0105038}}].

\bibitem{Hohm:2015ugy}
O.~Hohm and D.~Marques, {\it Perturbative double field theory on general
  backgrounds},  {\em Phys. Rev.} {\bf D93} (2016), no.~2 025032,
  [\href{http://arxiv.org/abs/1512.02658}{{\tt arXiv:1512.02658}}].

\bibitem{Hassler:2015pea}
F.~Hassler, {\em Double Field Theory on Group Manifolds (Thesis)}.
\newblock PhD thesis, Munich U., 2015.
\newblock \href{http://arxiv.org/abs/1509.07153}{{\tt arXiv:1509.07153}}.

\bibitem{Hohm:2010jy}
O.~Hohm, C.~Hull, and B.~Zwiebach, {\it {B}ackground independent action for
  double field theory},  {\em JHEP} {\bf 1007} (2010) 016,
  [\href{http://arxiv.org/abs/1003.5027}{{\tt arXiv:1003.5027}}].

\bibitem{Berman:2012vc}
D.~S. Berman, M.~Cederwall, A.~Kleinschmidt, and D.~C. Thompson, {\it {T}he
  gauge structure of generalised diffeomorphisms},  {\em JHEP} {\bf 01} (2013)
  064, [\href{http://arxiv.org/abs/1208.5884}{{\tt arXiv:1208.5884}}].

\bibitem{Hohm:2010xe}
O.~Hohm and S.~K. Kwak, {\it {F}rame-like {G}eometry of {D}ouble {F}ield
  {T}heory},  {\em J.Phys.} {\bf A44} (2011) 085404,
  [\href{http://arxiv.org/abs/1011.4101}{{\tt arXiv:1011.4101}}].

\bibitem{Grana:2008yw}
M.~Grana, R.~Minasian, M.~Petrini, and D.~Waldram, {\it T-duality, generalized
  geometry and non-geometric backgrounds},  {\em JHEP} {\bf 04} (2009) 075,
  [\href{http://arxiv.org/abs/0807.4527}{{\tt arXiv:0807.4527}}].

\bibitem{Lee:2014mla}
K.~Lee, C.~Strickland-Constable, and D.~Waldram, {\it {S}pheres, generalised
  parallelisability and consistent truncations},
  \href{http://arxiv.org/abs/1401.3360}{{\tt arXiv:1401.3360}}.

\bibitem{Dibitetto:2012rk}
G.~Dibitetto, J.~Fernandez-Melgarejo, D.~Marques, and D.~Roest, {\it {D}uality
  orbits of non-geometric fluxes},  {\em Fortsch.Phys.} {\bf 60} (2012)
  1123--1149, [\href{http://arxiv.org/abs/1203.6562}{{\tt arXiv:1203.6562}}].

\bibitem{Hull:2006va}
C.~M. Hull, {\it {D}oubled {G}eometry and {T}-{F}olds},  {\em JHEP} {\bf 0707}
  (2007) 080, [\href{http://arxiv.org/abs/hep-th/0605149}{{\tt
  hep-th/0605149}}].

\bibitem{Shelton:2006fd}
J.~Shelton, W.~Taylor, and B.~Wecht, {\it {G}eneralized {F}lux {V}acua},  {\em
  JHEP} {\bf 0702} (2007) 095, [\href{http://arxiv.org/abs/hep-th/0607015}{{\tt
  hep-th/0607015}}].

\bibitem{Plauschinn:2013wta}
E.~Plauschinn, {\it {T}-duality revisited},  {\em JHEP} {\bf 1401} (2014) 131,
  [\href{http://arxiv.org/abs/1310.4194}{{\tt arXiv:1310.4194}}].

\bibitem{Plauschinn:2014nha}
E.~Plauschinn, {\it On t-duality transformations for the three-sphere},  {\em
  Nucl. Phys.} {\bf B893} (2015) 257--286,
  [\href{http://arxiv.org/abs/1408.1715}{{\tt arXiv:1408.1715}}].

\bibitem{Hitchin:1999fh}
N.~J. Hitchin, {\it {L}ectures on special {L}agrangian submanifolds},
  \href{http://arxiv.org/abs/math/9907034}{{\tt math/9907034}}.

\bibitem{Lust:2010iy}
D.~Lüst, {\it {T}-duality and closed string non-commutative (doubled)
  geometry},  {\em JHEP} {\bf 1012} (2010) 084,
  [\href{http://arxiv.org/abs/1010.1361}{{\tt arXiv:1010.1361}}].

\bibitem{Andriot:2012an}
D.~Andriot, O.~Hohm, M.~Larfors, D.~Lüst, and P.~Patalong, {\it
  {N}on-{G}eometric {F}luxes in {S}upergravity and {D}ouble {F}ield {T}heory},
  {\em Fortsch.Phys.} {\bf 60} (2012) 1150--1186,
  [\href{http://arxiv.org/abs/1204.1979}{{\tt arXiv:1204.1979}}].

\bibitem{Blumenhagen:2012pc}
R.~Blumenhagen, A.~Deser, E.~Plauschinn, and F.~Rennecke, {\it Bianchi
  identities for non-geometric fluxes - from quasi-poisson structures to
  courant algebroids},  {\em Fortsch. Phys.} {\bf 60} (2012) 1217--1228,
  [\href{http://arxiv.org/abs/1205.1522}{{\tt arXiv:1205.1522}}].

\bibitem{Condeescu:2012sp}
C.~Condeescu, I.~Florakis, and D.~Lüst, {\it {A}symmetric {O}rbifolds,
  {N}on-{G}eometric {F}luxes and {N}on-{C}ommutativity in {C}losed {S}tring
  {T}heory},  {\em JHEP} {\bf 1204} (2012) 121,
  [\href{http://arxiv.org/abs/1202.6366}{{\tt arXiv:1202.6366}}].

\bibitem{Bakas:2015gia}
I.~Bakas and D.~Lüst, {\it {T}-duality, {Q}uotients and {C}urrents for
  {N}on-{G}eometric {C}losed {S}trings},  {\em Fortsch. Phys.} {\bf 63} (2015)
  543--570, [\href{http://arxiv.org/abs/1505.04004}{{\tt arXiv:1505.04004}}].

\bibitem{Mathai:2004qq}
V.~Mathai and J.~M. Rosenberg, {\it {T}-duality for torus bundles with
  ${H}$-fluxes via noncommutative topology},  {\em Commun. Math. Phys.} {\bf
  253} (2004) 705--721, [\href{http://arxiv.org/abs/hep-th/0401168}{{\tt
  hep-th/0401168}}].

\bibitem{Deser:2013pra}
A.~Deser, {\it {L}ie algebroids, non-associative structures and non-geometric
  fluxes},  {\em Fortsch. Phys.} {\bf 61} (2013) 1056--1153,
  [\href{http://arxiv.org/abs/1309.5792}{{\tt arXiv:1309.5792}}].

\bibitem{Hassler:2014sba}
F.~Hassler and D.~Lüst, {\it {C}onsistent {C}ompactification of {D}ouble
  {F}ield {T}heory on {N}on-geometric {F}lux {B}ackgrounds},  {\em JHEP} {\bf
  1405} (2014) 085, [\href{http://arxiv.org/abs/1401.5068}{{\tt
  arXiv:1401.5068}}].

\bibitem{deBoer:2012ma}
J.~de~Boer and M.~Shigemori, {\it {E}xotic {B}ranes in {S}tring {T}heory},
  {\em Phys.Rept.} {\bf 532} (2013) 65--118,
  [\href{http://arxiv.org/abs/1209.6056}{{\tt arXiv:1209.6056}}].

\bibitem{Hassler:2013wsa}
F.~Hassler and D.~Lüst, {\it {N}on-commutative/non-associative {IIA} ({IIB})
  {Q}- and {R}-branes and their intersections},  {\em JHEP} {\bf 1307} (2013)
  048, [\href{http://arxiv.org/abs/1303.1413}{{\tt arXiv:1303.1413}}].

\bibitem{Berkeley:2014nza}
J.~Berkeley, D.~S. Berman, and F.~J. Rudolph, {\it Strings and branes are
  waves},  {\em JHEP} {\bf 06} (2014) 006,
  [\href{http://arxiv.org/abs/1403.7198}{{\tt arXiv:1403.7198}}].

\bibitem{Berman:2014jsa}
D.~S. Berman and F.~J. Rudolph, {\it Branes are waves and monopoles},  {\em
  JHEP} {\bf 05} (2015) 015, [\href{http://arxiv.org/abs/1409.6314}{{\tt
  arXiv:1409.6314}}].

\bibitem{Bakhmatov:2016kfn}
I.~Bakhmatov, A.~Kleinschmidt, and E.~T. Musaev, {\it Non-geometric branes are
  dft monopoles},  {\em JHEP} {\bf 10} (2016) 076,
  [\href{http://arxiv.org/abs/1607.05450}{{\tt arXiv:1607.05450}}].

\bibitem{Deser:2014mxa}
A.~Deser and J.~Stasheff, {\it Even symplectic supermanifolds and double field
  theory},  {\em Commun. Math. Phys.} {\bf 339} (2015), no.~3 1003--1020,
  [\href{http://arxiv.org/abs/1406.3601}{{\tt arXiv:1406.3601}}].

\bibitem{Deser:2016qkw}
A.~Deser and C.~Saemann, {\it Extended riemannian geometry i: Local double
  field theory},  \href{http://arxiv.org/abs/1611.02772}{{\tt
  arXiv:1611.02772}}.

\bibitem{Blumenhagen:2010hj}
R.~Blumenhagen and E.~Plauschinn, {\it {N}onassociative {G}ravity in {S}tring
  {T}heory?},  {\em J.Phys.} {\bf A44} (2011) 015401,
  [\href{http://arxiv.org/abs/1010.1263}{{\tt arXiv:1010.1263}}].

\bibitem{Blumenhagen:2011ph}
R.~Blumenhagen, A.~Deser, D.~Lüst, E.~Plauschinn, and F.~Rennecke, {\it
  {N}on-geometric {F}luxes, {A}symmetric {S}trings and {N}onassociative
  {G}eometry},  {\em J.Phys.} {\bf A44} (2011) 385401,
  [\href{http://arxiv.org/abs/1106.0316}{{\tt arXiv:1106.0316}}].

\bibitem{Mylonas:2012pg}
D.~Mylonas, P.~Schupp, and R.~J. Szabo, {\it {M}embrane {S}igma-{M}odels and
  {Q}uantization of {N}on-{G}eometric {F}lux {B}ackgrounds},  {\em JHEP} {\bf
  1209} (2012) 012, [\href{http://arxiv.org/abs/1207.0926}{{\tt
  arXiv:1207.0926}}].

\bibitem{Bakas:2013jwa}
I.~Bakas and D.~Lüst, {\it 3-{C}ocycles, {N}on-{A}ssociative {S}tar-{P}roducts
  and the {M}agnetic {P}aradigm of ${R}$-{F}lux {S}tring {V}acua},  {\em JHEP}
  {\bf 1401} (2014) 171, [\href{http://arxiv.org/abs/1309.3172}{{\tt
  arXiv:1309.3172}}].

\bibitem{Blumenhagen:2013zpa}
R.~Blumenhagen, M.~Fuchs, F.~Hassler, D.~Lüst, and R.~Sun, {\it
  {N}on-associative {D}eformations of {G}eometry in {D}ouble {F}ield {T}heory},
   {\em JHEP} {\bf 1404} (2014) 141,
  [\href{http://arxiv.org/abs/1312.0719}{{\tt arXiv:1312.0719}}].

\bibitem{Ramgoolam:2001zx}
S.~Ramgoolam, {\it {O}n spherical harmonics for fuzzy spheres in diverse
  dimensions},  {\em Nucl.Phys.} {\bf B610} (2001) 461--488,
  [\href{http://arxiv.org/abs/hep-th/0105006}{{\tt hep-th/0105006}}].

\end{thebibliography}\endgroup
\bibliographystyle{JHEP}
\end{document}